\documentclass[10pt,letterpaper]{article}
\usepackage[top=0.85in,left=2.75in,footskip=0.75in]{geometry}

\usepackage{amsmath,amssymb}

\usepackage{changepage}

\usepackage{textcomp,marvosym}

\usepackage{cite}

\usepackage{nameref,hyperref}

\usepackage[right]{lineno}

\usepackage[nopatch=eqnum]{microtype}
\DisableLigatures[f]{encoding = *, family = * }

\usepackage[table]{xcolor}

\usepackage{array}

\newcolumntype{+}{!{\vrule width 2pt}}

\newlength\savedwidth

\usepackage{setspace} 

\raggedright
\usepackage[aboveskip=1pt,labelfont=bf,labelsep=period,justification=raggedright,singlelinecheck=off]{caption}

\makeatletter
\renewcommand{\@biblabel}[1]{\quad#1.}
\makeatother

\usepackage{lastpage,fancyhdr,graphicx}
\usepackage{epstopdf}
\fancyheadoffset[L]{2.25in}
\fancyfootoffset[L]{2.25in}
\usepackage{siunitx}
\usepackage{todonotes}
\reversemarginpar
\usepackage{csquotes}
\usepackage{float}
\usepackage{caption}
\usepackage{subcaption}
\usepackage{booktabs}

\newcommand{\leak}{{\text{leak}}}

\newcommand{\Vth}{{V_\text{th}}}
\newcommand{\ext}{{\text{ext}}}

\newcommand{\f}{\text{f}}
\newcommand{\s}{\text{s}}
\newcommand{\us}{\text{u}}

\DeclareSIUnit{\mScmsq}{\milli\siemens\per\centi\meter\squared} 
\DeclareSIUnit\Molar{\textsc{M}}
\DeclareSIUnit{\mM}{\milli\Molar}
\DeclareSIUnit{\uM}{\micro\Molar}

\begin{document}
\vspace*{0.2in}

\begin{flushleft}
{\Large
\textbf{Fast reconstruction of degenerate populations of conductance-based neuron models from spike times} 
}
\newline
\\
Julien Brandoit\textsuperscript{1},
Damien Ernst\textsuperscript{1,2},
Guillaume Drion\textsuperscript{1\Yinyang},
Arthur Fyon\textsuperscript{1\Yinyang*}
\\
\bigskip
\textbf{1} Montefiore Institute, University of Liège, 10 Allée de la Découverte, Liège, 4000, Belgium
\\
\textbf{2} LTCI, Telecom Paris, Institut Polytechnique de Paris, 19 Place Marguerite Perey, Palaiseau, 91120, France 
\\
\bigskip

%
%
\Yinyang These authors contributed equally to this work.





* afyon@uliege.be

\end{flushleft}


%
\section*{Abstract}
Inferring the biophysical parameters of conductance-based models (CBMs) from experimentally accessible recordings remains a central challenge in computational neuroscience. Spike times are the most widely available data, yet they reveal little about which combinations of ion channel conductances generate the observed activity. This inverse problem is further complicated by neuronal degeneracy, where multiple distinct conductance sets yield similar spiking patterns. We introduce a method that addresses this challenge by combining deep learning with Dynamic Input Conductances (DICs), a theoretical framework that reduces complex CBMs to three interpretable feedback components governing excitability and firing patterns. Our approach first maps spike times directly to DIC densities at threshold using a lightweight neural network that learns a low-dimensional representation of neuronal activity. The predicted DIC values are then used to generate degenerate CBM populations via an iterative compensation algorithm, ensuring compatibility with the intermediate target DICs, and thereby reproducing the corresponding firing patterns, even in high-dimensional models. Applied to two neuronal models, this algorithmic pipeline reconstructs spiking, bursting, and irregular regimes with high accuracy and robustness to variability, including spike trains generated under noisy current injection mimicking physiological stochasticity. It produces diverse degenerate populations within milliseconds on standard hardware, enabling scalable and efficient inference from spike recordings alone. Beyond methodological advances, we provide an open-source software package with a graphical interface that allows experimentalists to generate and explore CBM populations directly from spike trains without requiring programming expertise. Together, this work positions DICs as a practical and interpretable link between experimentally observed activity and mechanistic models. By enabling fast and scalable reconstruction of degenerate populations directly from spike times, our approach provides a powerful way to investigate how neurons exploit conductance variability to achieve reliable computation and provides the foundation for experimental applications that span from neuromodulation studies to real-time model-guided interventions.


%
\section*{Author summary}
Neurons communicate through spikes, and spike timing is a crucial part of neuronal processing. Spike times can be recorded experimentally both intracellularly and extracellularly, and are the main output of state-of-the-art neural probes. On the other hand, neuronal activity is controlled at the molecular level by the currents generated by many different transmembrane proteins called ion channels. Connecting spike timing to ion channel composition remains an arduous task to date. To address this challenge, we developed a method that combines deep learning with a theoretical tool called Dynamic Input Conductances (DICs), which reduce the complexity of ion channel interactions into three interpretable components describing how neurons spike. Our approach uses deep learning to infer DICs directly from spike times and then generates populations of \enquote{twin} neuron models that replicate the observed activity while capturing natural variability in membrane channel composition. The method is fast, accurate, and works using only spike recordings. We also provide open-source software with a graphical interface, making it accessible to researchers without programming expertise.
\nolinenumbers

\section*{Introduction}

A central objective in both experimental and computational neuroscience is to understand how neurons generate and regulate electrical activity from their ion channels, linking microscopic mechanisms to macroscopic dynamics. Extracellular recordings, particularly spike times, remain the most widely accessible data, especially when probing large networks \cite{tiesingaRegulationSpikeTiming2008, mainenReliabilitySpikeTiming1995, williamsDiscoveringPreciseTemporal2020}. Spike times provide rich information about neural dynamics across brain regions and conditions. However, while they reveal what neurons do and how they respond to external perturbations, they rarely reveal the underlying mechanisms, specifically which combinations of biophysical parameters produce the observed activity. This question is crucial because such biophysical parameters are, among others, targets of neuromodulators and neuroactive medications \cite{marderNeuromodulationNeuronalCircuits2012, bargmannConnectomeBrainFunction2013, mccormickNeuromodulationBrainState2020}.

A major obstacle in linking neuronal activity with biophysical properties is \emph{neuronal degeneracy}, the well-established fact that multiple distinct parameter sets, particularly effective ion channel densities, can produce similar spiking behaviors \cite{goaillardIonChannelDegeneracy2021}. This functional flexibility enhances robustness but complicates interpretation, as neurons exhibit inherently nonlinear, high-dimensional dynamics. When different parameter combinations produce indistinguishable activity, identifying which configurations underlie recorded behavior becomes an underdetermined problem \cite{goldmanGlobalStructureRobustness2001a, liuModelNeuronActivityDependent1998, golowaschActivityDependentRegulationPotassium1999a}.

Conductance-based models (CBMs) provide a principled framework to tackle this challenge computationally. Their parameters, such as maximal ion conductances, link directly to measurable biological properties like ion channel expression. However, realistic CBMs are complex, often requiring dozens of parameters \cite{liuModelNeuronActivityDependent1998, taylorHowMultipleConductances2009, yuMathematicalModelMidbrain2015, fyonNewPerspective2026}. Combined with degeneracy, this complexity makes inferring conductances from experimental data highly challenging and often computationally intractable, with results that are difficult to interpret.

This difficulty defines a core inverse problem in computational neuroscience: \emph{given an observed activity pattern such as a spike train, how can one identify a set (or better, the population) of biophysical models that could have generated it?} Existing approaches face key limitations. Many require full voltage traces or intracellular recordings, making them sensitive to noise and less feasible in practice \cite{druckmannNovelMultipleObjective2007, prinzAlternativeHandTuningConductanceBased2003, shepardsonAlgorithmsInvertingHodgkinHuxley2009, huysEfficientEstimationDetailed2006, burghiFeedbackIdentificationConductancebased2021, burghiDistributedOnlineEstimation2022, mengSequentialMonteCarlo2011, mengUnifiedApproachLinking2014, ditlevsenEstimationPartiallyObserved2014}. Recent advances in simulation-based inference (SBI) have partially addressed these limitations by training neural density estimators to approximate posterior distributions over parameters given observed data, enabling fast inference while capturing parameter uncertainty~\cite{goncalvesTrainingDeepNeural2020, cranmerFrontierSimulationbasedInference2020, lueckmannFlexibleStatisticalInference2017, vetter2024sourcerersamplebasedmaximumentropy, burghiRapidInterpretable2025, burghiQuantitativePrediction2025}. Related efforts using data assimilation and variational methods have a long history in this context, notably the work of Abarbanel and collaborators on dynamical state and parameter estimation in  conductance-based models~\cite{abarbanel2009dynamical, meliza2014estimating}, as well as subsequent  applications to automated model construction~\cite{nogaretAutomaticConstructionPredictive2016}. However, these approaches operate directly in the high-dimensional conductance space without explicit reference to the dynamical mechanisms that shape neuronal activity. As a result, the learned posteriors, though valid, do not straightforwardly reveal \emph{why} certain parameter combinations produce similar outputs, and the relationship between parameter variability and functional equivalence remains implicit.

In this work, we introduce a method that addresses this inverse problem by combining deep learning with Dynamic Input Conductances (DICs)~\cite{drionDynamicInputConductances2015, fyonDimensionalityReductionNeuronal2024}, a theoretical framework from dynamical systems theory that reduces complex CBMs to three interpretable feedback components governing excitability and firing patterns. Our approach uses DICs as a low-dimensional intermediate: a lightweight deep learning architecture first maps spike times to DIC values at threshold, and an iterative compensation algorithm then generates degenerate CBM populations compatible with these values. Given only spike times, the method outputs a population of degenerate CBMs within milliseconds on standard hardware.

We validate the pipeline on two distinct models and show that it faithfully reconstructs neuronal activity across spiking, bursting, and irregular regimes, while maintaining robustness to variability and noise. By combining deep learning with DIC theory, this approach provides a practical solution to the inverse inference problem. It bridges experimentally accessible observables and mechanistic CBMs, demonstrates that DICs serve as interpretable low-dimensional intermediates, and enables scalable reconstruction of degenerate populations from spike times alone.

To support widespread adoption and facilitate daily use in experimental settings, we provide our entire pipeline as an open-source software package with a graphical interface \cite{brandoit2025spike2pop}. Recognizing that many researchers in the biomedical sciences come from diverse backgrounds where programming and numerical modeling are not core skills, the tool is designed to be intuitive and easy to use, allowing experimentalists to generate and explore model populations directly from spike recordings without writing code.

\section*{Results}

\subsection*{General problem statement}
In this work, we assumed a known conductance-based model (CBM) whose membrane dynamics is described by:
\begin{equation*}
    C \frac{dV}{dt} + g_\text{leak}(V - E_\text{leak}) 
    = - \sum_{i \in \mathcal{I}} \bar{g}_{i} m_{i}^{p_{i}} h_{i}^{q_{i}} (V - E_{i}) + I_{\text{ext}}\quad,
\end{equation*}
where $V$ denotes the membrane potential, $C$ is the membrane capacitance, $g_\text{leak}$ and $E_\text{leak}$ are the leak conductance and reversal potential, and each ionic current $i \in \mathcal{I}$ is characterized by its maximal conductance $\bar{g}_i$, gating variables $m_i$ and $h_i$ with powers $p_i$ and $q_i$, and reversal potential $E_i$. The external current input is denoted by $I_{\text{ext}}$. 
In this formulation, the structure of the model (i.e., the functional form of the ionic currents and their gating dynamics) was assumed to be fixed and known (see Materials and Methods). The unknown parameters are the conductances $\bar{g} = [\bar{g}_1, \dots, \bar{g}_{|\mathcal{I}|}, g_\leak] \in \mathcal{G}$, which must be inferred from observations extracted from the voltage trace. A specific choice of $\bar{g}$ fully specifies one instance of the considered CBM, while all other parameters (e.g., reversal potentials, capacitance, \dots) are assumed to be known and fixed; inferring these remains an avenue for future work. We chose spike times as representation of the activity, as these data are easily accessible from both intracellular and extracellular recordings. We denote such a recorded spike times sequence by:
\begin{equation*}
    x = [t_1, t_2, \dots, t_{N_\text{spikes}}]\quad.
\end{equation*}
The quantity $N_\text{spikes}$ denotes the total number of spikes detected in the recording. Its value depends both on the duration of the recording and on the firing activity of the neuron, which makes $x$ a variable-length representation of the neuronal activity.

Due to degeneracy \cite{marderMultipleModelsCapture2011, marderNeuromodulationNeuronalCircuits2012}, the mapping:  
\begin{equation*} 
    x \longmapsto \bar{g} \in \mathcal{G}^*(x)\quad,
\end{equation*}
is not bijective: the solution to the inference problem is not a single point, but rather a subspace $\mathcal{G}^*(x) \subset \mathcal{G}$ containing infinitely many parameter sets compatible with $x$ \cite{fyonDimensionalityReductionNeuronal2024, drionDynamicInputConductances2015, golowaschFailureAveragingConstruction2002, vetter2024sourcerersamplebasedmaximumentropy}.

Our objective is to build a set of $P$ models (with $P$ freely chosen by the experimentalist) with different conductance values that all reproduce a firing pattern similar to $x$. We call this set $\mathcal{P}$, and since each model is defined by its value of $\bar{g}$ we write $\mathcal{P} = \left\{[\bar{g}_1, \dots, \bar{g}_{|\mathcal{I}|}, g_\leak]_i\right\}_{i=1}^P$. We can build such a set by generating $P$ instances from the subspace $\mathcal{G}^*(x)$. 


While existing methods either infer a single solution or attempt to learn the full high-dimensional solution space directly (see Introduction), we adopt an intermediate strategy. This strategy divides the problem into two parts and leverages intermediate low-dimensional representations of CBMs called DICs (see Materials and Methods) \cite{drionDynamicInputConductances2015, fyonDimensionalityReductionNeuronal2024, fyonReliableNeuromodulationAdaptive2023, fyon2024neuromodulation}. For any CBM, we can analytically construct a DIC representation determined by:
\begin{equation}\label{eq:dics_mapping}
    g_\text{DICs}(V) = S(V; \bar{g}) \cdot \bar{g}\quad,
\end{equation}
where $S = \frac{\partial g_{\text{DICs}}}{\partial \bar{g}},$ is the sensitivity matrix determined by the CBM as in~\cite{fyonDimensionalityReductionNeuronal2024}, and $g_\text{DICs}(V) : V \mapsto \mathbb{R}^3$ provides a low-dimensional equivalent of the high-dimensional conductance vector $\bar{g}$. Importantly, the DIC curves evaluated at a specific potential, the threshold potential $\Vth$, are sufficient to capture most of the spontaneous activity associated with a given CBM~\cite{fyonDimensionalityReductionNeuronal2024, drionDynamicInputConductances2015}. The threshold potential corresponds to the voltage at which the neuron is maximally sensitive to changes in its conductance parameters (see Materials and Methods).


The three components of the DIC representation correspond to distinct timescales of the membrane dynamics, giving rise to three separate DIC curves: the fast component $g_\text{f}(V)$, the slow component $g_\text{s}(V)$, and the ultra-slow component $g_\text{u}(V)$. Each curve captures the influence of ionic conductances acting on its respective timescale. Because these curves evaluated at a single characteristic voltage are sufficient to capture most of the spontaneous activity (Eq.~\ref{eq:dics_mapping}), the full conductance space effectively reduces to just three scalars, providing a compact summary of all conductance configurations that produce similar firing activity.

\begin{figure}[H]
  \centering
  \begin{adjustwidth}{-0in}{0in}
  \includegraphics[width=5.2in]{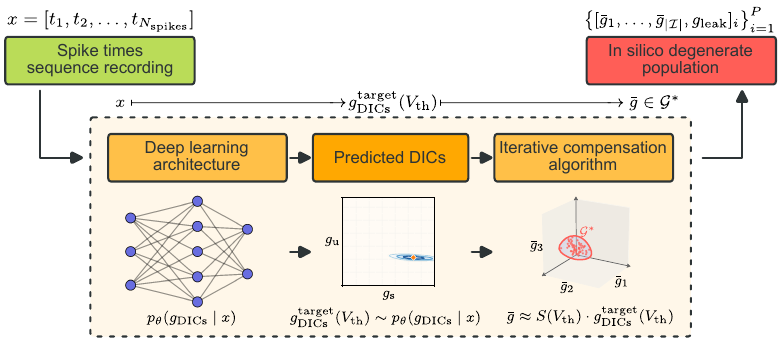}
  \caption{\textbf{Our proposed approach.}
  Spike time sequences are processed by a deep learning model that predicts DICs, a compact representation of the high-dimensional conductance space. These predicted DIC values serve as targets for an iterative compensation algorithm, which explores the degenerate solution space to generate multiple conductance configurations $\bar{g}$ that reproduce the input spike pattern. This two-step strategy (i) reduces the dimensionality of the inference problem, and (ii) leverages compensation to recover diverse biologically plausible parameter sets consistent with the observed activity.}
  \label{fig:fig_pipeline_intro}
  \end{adjustwidth}
\end{figure}

Leveraging this intermediate representation, our approach proceeds in two steps (Fig~\ref{fig:fig_pipeline_intro}). First, a \emph{deep learning architecture} learns the mapping from observed spike times $x$ to the corresponding DIC values. Because different activity regimes are governed by a different number of timescales, degeneracy can also arise within the DIC space itself, particularly for spiking activity; we therefore perform posterior density estimation and sample from the learned density to obtain $g_\text{DICs}^\text{target}(\Vth)$. This enables parameter inference in a tractable, low-dimensional space by inferring the left-hand side of relation~(\ref{eq:dics_mapping}). Second, given a predicted DIC vector, we generate valid CBM conductance vectors $\bar{g}$ that are compatible with these DIC values, thus recovering diverse, biologically plausible instances of the original high-dimensional model. For this step, we build upon the compensation method introduced in~\cite{fyonDimensionalityReductionNeuronal2024}, which generates degenerate CBM populations from DIC constraints, employing an iterative extension that improves constraint satisfaction in models with nonlinear compensatory structure (see Materials and Methods). Overall, our approach reads:
\begin{equation*}
    x \xrightarrow[]{\text{Deep learning architecture}} g_\text{DICs}^\text{target}(\Vth)\xrightarrow[]{\text{Compensation algorithm}}\bar{g}\in \mathcal{G}^*\quad.
\end{equation*}

This work addresses three components: (i) building a synthetic dataset linking DIC values to a diverse range of neuronal activities; (ii) designing a deep learning architecture capable of processing variable-length spike time sequences and predicting plausible DIC targets; and (iii) validating the complete pipeline, from spike times to degenerate populations, across models and under realistic noise conditions.

\subsection*{DICs provide a structured and learnable representation of neuronal activity}

To train the deep learning architecture and characterize the relationship between DIC values and neuronal activity, we constructed a large synthetic dataset spanning a broad range of DIC values and corresponding spike trains \cite{brandoit2025degenerate}. Rather than sampling conductance parameters directly, we uniformly sampled the slow and ultra-slow DICs $g^* = \left(g_\s(\Vth);~g_\us(\Vth)\right)$ across a wide bounded region, and generated for each sampled pair a degenerate population of CBM instances using the iterative compensation algorithm. Each instance was simulated under noisy current injection mimicking physiological stochasticity, yielding a total of $|\mathcal{T}| = 1{,}200{,}000$ neurons spanning spiking, bursting, and irregular regimes. Full details of the dataset construction are provided in the Materials and Methods.

\begin{figure}[H]
  \centering
  \begin{adjustwidth}{-2.3in}{0in}
  \includegraphics[width=7.5in]{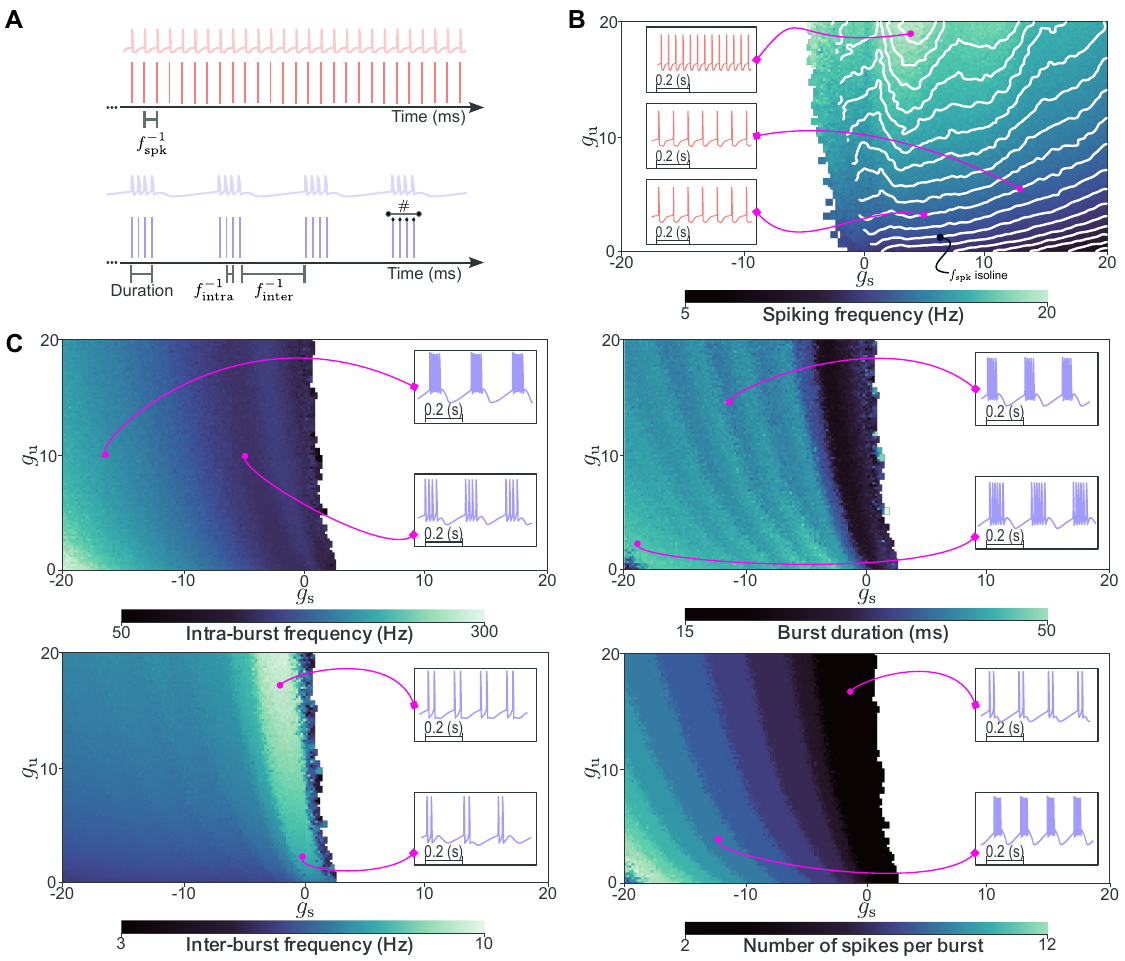}
    \caption{{\bf Activity descriptors vary smoothly across the DIC space.}
    Heatmaps show how individual activity metrics vary across the DIC space, revealing clear gradients that reflect a nonlinear yet structured relationship between DIC constraints and neuronal firing patterns.
    \textbf{(A)} Representative example traces of a spiking neuron (top) and a bursting neuron (bottom), with annotations highlighting the descriptors extracted from each regime: mean spiking frequency for the spiking case, and mean burst duration, mean intra-burst frequency, mean inter-burst frequency, and mean number of spikes per burst for the bursting case.
    \textbf{(B)} Spiking metrics summarized by mean firing rate.
    \textbf{(C)} Bursting metrics including intra-burst frequency, inter-burst frequency, burst duration, and spikes per burst.
    These smooth gradients support the learnability of the inverse mapping from spike trains to DICs, and position DICs as meaningful low-dimensional intermediates for characterizing neuronal activity.}
    \label{fig:fig_dics_features}
    \end{adjustwidth}
\end{figure}

This dataset reveals a clear and structured relationship between DIC values and neuronal firing patterns (Fig~\ref{fig:fig_dics_features}). Each activity descriptor can be viewed as a function of the sampled slow and ultra-slow DICs. A sharp separation between spiking and bursting neurons emerges, with a transition largely independent of $g_\us$ and located near $g_\s \approx 0$, consistent with prior work \cite{fyonDimensionalityReductionNeuronal2024, drionDynamicInputConductances2015}. Negative slow DIC values at threshold are associated with bursting, while positive values lead to spiking. The overlapping zone at the transition is due to heterogeneous populations. Outside this narrow transition zone, generated populations are fully homogeneous: all instances exhibit the nominal activity type (spiking or bursting) with no failures, as quantified by the entropy measure reported in the Supporting Information (see \nameref{S1_appendix}).

Beyond this qualitative separation, smooth gradients appear across the DIC space for both bursting descriptors (Fig~\ref{fig:fig_dics_features}A) and spiking frequency (Fig~\ref{fig:fig_dics_features}B), revealing a structured and nonlinear relationship between DIC targets and resulting activity. The ranges of the colorbars confirm that the dataset spans a wide variety of activity descriptors (similar to those reported in \cite{prinzAlternativeHandTuningConductanceBased2003}). Crucially, different neuronal activities are characterized by different DIC values, yet multiple distinct DIC configurations can map to the same activity regime, especially in spiking. This many-to-one correspondence reflects the intrinsic degeneracy of neuronal systems and the timescale separation between activity regimes.

This is the first dataset to systematically map descriptive activity metrics across the DIC space. The smoothness of these gradients is particularly important: it implies that small changes in DIC values correspond to gradual changes in firing behavior, which supports the learnability of the inverse mapping from spike times to DICs by a neural network. Together, these patterns establish DICs as a practical, low-dimensional, and interpretable intermediate that quantitatively links conductance configurations to experimentally observable firing patterns.

The utility of DICs as a learnable intermediate, however, depends on the ability to map predicted DIC values back to conductance vectors $\bar{g}$. This inverse step exploits the relationship $g_\text{DICs}(\Vth) = S(\Vth; \bar{g}) \cdot \bar{g}$ between DICs and maximal conductances. By partitioning $\bar{g}$ into a randomly sampled subset $\bar{g}_\text{random}$ and a compensated subset $\bar{g}_\text{comp.}$, the system can be solved for $\bar{g}_\text{comp.}$ given a target DIC vector, while different draws of $\bar{g}_\text{random}$ yield different valid solutions, thereby generating a degenerate population as in~\cite{fyonDimensionalityReductionNeuronal2024, fyon2024neuromodulation}. We emphasize that the partition is exhaustive: together, the two subsets contain all components of $\bar{g}$. This single-step linear compensation is exact when the sensitivity matrix $S$ does not depend on the compensated conductances. However, in models where internal dynamics (such as intracellular calcium concentration in the STG model) introduce nonlinear dependencies, the linear approximation can leave residual errors between target and enforced DIC values. To address this, we employ an iterative extension in which the sensitivity matrix is recomputed at each iteration based on the current estimate of $\bar{g}_\text{comp.}$, progressively reducing these residuals (see Materials and Methods). In practice, five iterations suffice to reduce residual norms by approximately a factor of 15 compared to the single-step method, yielding tighter distributions of activity statistics across generated populations while preserving degeneracy (see Supporting Information, \nameref{S1_appendix}). This compensation step is used both for generating the synthetic training dataset and as an integral part of the full pipeline at inference time.

\subsection*{A deep learning architecture efficiently encodes spike trains into DIC targets}

The first block of our pipeline (Fig~\ref{fig:fig_pipeline_intro}) is a deep learning architecture that maps raw spike time sequences directly to a density over DIC values at threshold, without relying on hand-crafted summary statistics. From this learned density, a plausible DIC target is sampled and passed to the iterative compensation algorithm. The architecture and training procedure are described in detail in the Materials and Methods.

To assess whether the learned latent representation captures biologically meaningful structure, we evaluated two auxiliary prediction heads operating solely on the latent embedding: a classification head predicting firing regime (spiking or bursting), and a regression head predicting key activity descriptors. To isolate intrinsic properties from input-driven variability, these metrics were computed on an equivalent dataset simulated without injected current, such that activity descriptors reflect solely the intrinsic conductance properties of each neuron, free from the additional variability introduced by injected stochastic inputs. The model achieved $99.83\%$ classification accuracy and accurately predicted mean spike frequency, intra- and inter-burst frequencies, burst duration, and number of spikes per burst, with mean absolute errors substantially lower than the intrinsic variability of the dataset (Table~\ref{table:best_model_mae_std}). This confirms that the encoder captures task-relevant temporal structure from raw spike trains.

\begin{table}[htbp]
  \centering
  \caption{\textbf{Auxiliary regression performance on the STG neuron dataset.} Mean absolute error (MAE) values for auxiliary regression tasks are substantially lower than the inherent dataset variability (standard deviation). Descriptors include mean spike frequency ($f_\text{spk}$), intra- and inter-burst frequencies ($f_\text{intra}$, $f_\text{inter}$), burst duration, and mean number of spikes per burst ($\#$).}
  \label{table:best_model_mae_std}
  \renewcommand{\arraystretch}{0.8}
  \begin{tabular}{lccccc}
    \toprule
    Descriptor & $f_\text{spk}$ & $f_\text{intra}$ & $f_\text{inter}$ & Duration & $\#$ \\
    \midrule
    MAE & 0.11~Hz & 2.26~Hz & 0.15~Hz & 0.81~ms & 0.16 \\
    Dataset std & 3.02~Hz & 47.91~Hz & 1.11~Hz & 7.85~ms & 2.25 \\
    \bottomrule
  \end{tabular}
\end{table}

Because degeneracy manifests within the DIC space itself, particularly in the spiking regime where multiple DIC configurations yield functionally indistinguishable activity, pointwise error metrics on predicted DIC values are not a meaningful proxy for pipeline performance. The ultimate criterion is whether the generated populations faithfully reproduce the activity of the input spike trains, which we assess in the following section through a systematic comparison of activity descriptors between input and generated populations. We provide in \nameref{S1_appendix} evidence that the learned posterior densities are well calibrated following recommendations from~\cite{hermans2022trustcrisissimulationbasedinference}, demonstrating that the architecture correctly captures degeneracy in the DIC space across both spiking and bursting regimes.

Inference is highly efficient, and, combined with the iterative compensation algorithm, the full pipeline produces degenerate CBM populations within milliseconds on standard hardware.

\subsection*{The complete generative pipeline reconstructs accurate and diverse degenerate populations from spike times}

The complete pipeline combines the deep learning architecture with the iterative compensation algorithm to transform spike time sequences into \textit{in silico} degenerate populations of CBMs, using DICs at threshold as intermediates (Fig~\ref{fig:fig_pipeline_intro}). This approach generates populations rather than single solutions, capturing a diverse set of conductance combinations compatible with the input activity.

Concretely, given a spike train, the deep learning architecture encodes its temporal structure and yields a density over DIC values compatible with the observed firing pattern, from which one plausible DIC target is sampled. This target is then provided to the iterative compensation algorithm, which generates a set of maximal conductance configurations satisfying the DIC constraints. Because all generated conductance sets are anchored to the same DIC target, they are by construction associated with activity that reproduces the input firing pattern, while spanning a broad and diverse region of the conductance space.

\begin{figure}[H]
  \centering
  \begin{adjustwidth}{-0in}{0in}
  \includegraphics[width=5.2in]{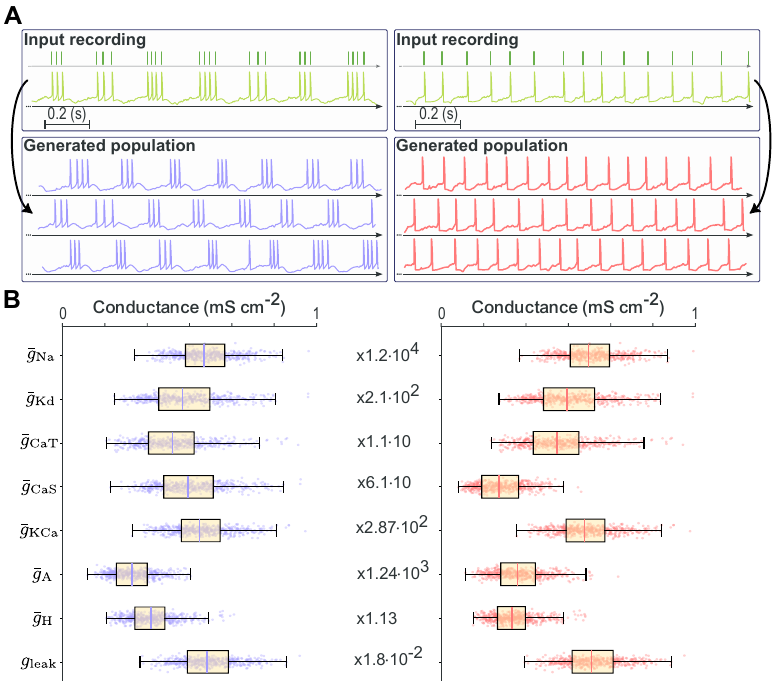}
  \caption{\textbf{Backbone pipeline output for the STG neuron model.}
  \textbf{(A)} Target spike trains (dark green, top) compared with three representative generated spike trains for spiking (red, right) and bursting (purple, left) regimes, showing accurate reproduction of activity patterns.
  \textbf{(B)} Distributions of maximal conductances across 500 generated neurons in spiking (red, right) and bursting (purple, left) regimes, demonstrating broad parameter variability despite similar dynamics.}
  \label{fig:fig_pipeline}
  \end{adjustwidth}
\end{figure}

Qualitative inspection confirms close agreement between generated and target activity patterns in both spiking and bursting regimes (Fig~\ref{fig:fig_pipeline}A). For each regime, a representative input spike train (dark green) is shown alongside three generated traces, illustrating that the pipeline produces outputs whose temporal structure closely matches the input, with variability confined to levels expected under degeneracy and noisy injected current. Importantly, the method operates solely on spike time input, without requiring voltage traces. However, because the method is blind to subthreshold dynamics, certain features such as after-depolarizations shaping the interspike interval cannot be reproduced.

Across 500 generated neurons per regime, maximal conductance distributions are broad yet yield equivalent spiking or bursting dynamics (Fig~\ref{fig:fig_pipeline}B), confirming that the pipeline reconstructs multiple distinct solutions producing similar functional outputs, consistent with population-level degeneracy.

\begin{figure}[H]
  \centering
  \begin{adjustwidth}{-0in}{0in}
  \includegraphics[width=5.2in]{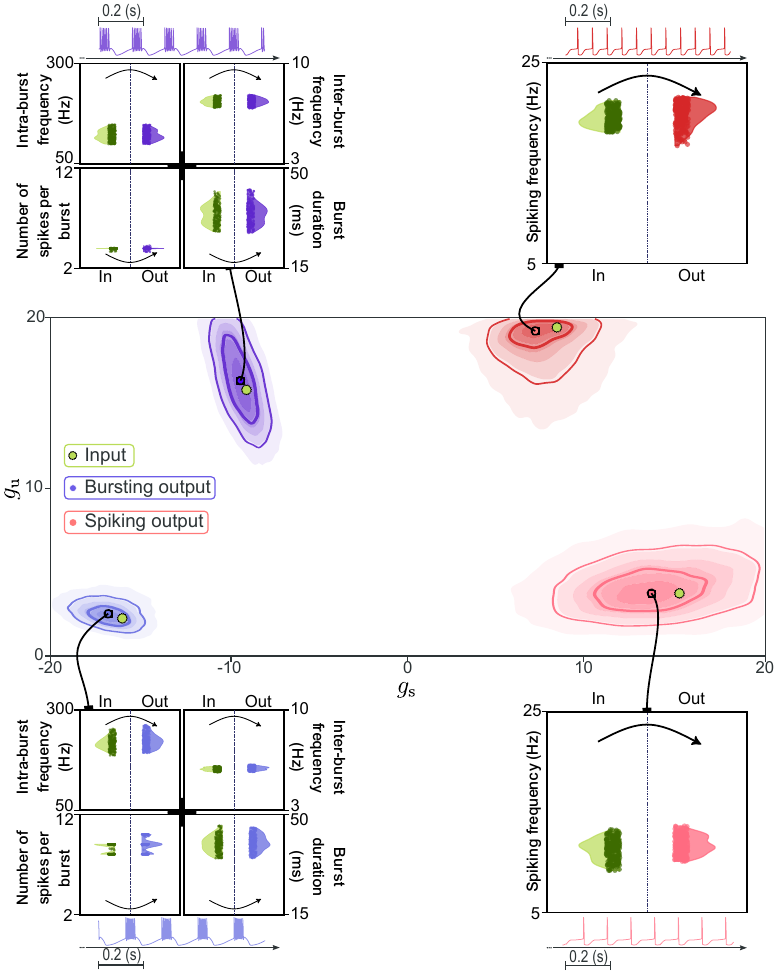}
  \caption{\textbf{Quantitative comparison of input and generated populations.}
  Four target DIC values (green dots) are selected in the $(g_\s, g_\us)$ plane, spanning spiking and bursting regimes. For each target, an input population of $256$ degenerate neurons is generated via iterative compensation. Spike times from each input neuron are passed through the full pipeline: the deep learning architecture infers a conditional density over DICs (shown as contour plots), from which one pair of DIC targets is sampled and used to generate one neuron each via iterative compensation. Violin plots compare distributions of key activity metrics between the input population (left, green) and the output population (right, colored) for each target point, demonstrating faithful reproduction of firing statistics across regimes.}
  \label{fig:fig_quantitative}
  \end{adjustwidth}
\end{figure}

To quantitatively evaluate reconstruction accuracy, we designed an end-to-end test that mirrors how the pipeline could be used in practice (Fig~\ref{fig:fig_quantitative}). We selected four target DIC values in the $(g_\s, g_\us)$ plane, spanning both the spiking and bursting regimes. For each target, we generated an \textit{input population} of $256$ neurons using the iterative compensation algorithm. These neurons share identical DIC values at threshold and are therefore degenerate by construction, but still display small variability in their firing patterns arising from three sources: the approximation inherent in constraining DICs only at threshold, the iterative compensation procedure, and the noisy injected current. Such variability is compatible with degeneracy and reflects biological heterogeneity. This first step would, in practice, be replaced by an experimental recording of the activity of a neuron.

We then passed the spike times of each neuron in the input population through the full pipeline. For each input spike train, the deep learning architecture produces a conditional density over DIC values, from which one pair of plausible DIC targets is sampled. Each sampled DIC target is in turn used by the iterative compensation algorithm to generate one new neuron, yielding a total of 256 neurons per target point, which collectively form the \textit{output population}.

Fig~\ref{fig:fig_quantitative} shows the inferred DIC densities for each of the four target points, alongside violin plots comparing key activity descriptors between input and output populations. In the bursting regime, the inferred densities are concentrated around the target DIC values, and the output populations closely reproduce the input distributions of intra-burst frequency, inter-burst frequency, burst duration, and number of spikes per burst for both input examples. In the spiking regime, the inferred densities spread more broadly compared to bursting ones, due to the one-dimensional manifold identified earlier (Fig~\ref{fig:fig_dics_features}B), reflecting the intrinsic DIC degeneracy of spiking activity where multiple $(g_\s, g_\us)$ combinations yield similar firing rates. Despite this spread in DIC space, the output spiking frequency distributions remain well matched to the inputs, confirming that the pipeline correctly handles this ambiguity by sampling from a region of DIC space that is functionally equivalent. Overall, these results demonstrate that the pipeline faithfully recovers activity features across regimes while preserving degeneracy.

\subsection*{The pipeline extends to new conductance-based models with minimal retraining}

To demonstrate the scalability of our pipeline across distinct CBMs, we applied low-rank adaptation (LoRA) \cite{huLoRALowRankAdaptation2021} to transfer the backbone model trained on the STG dataset to a dopaminergic (DA) neuron model (see Materials and Methods). This setting highlights a drastic change of context: from a bursting neuron of the stomatogastric ganglion, embedded in the gastric mill rhythm, to a slow pacemaker DA neuron with a completely different conductance composition and dynamical repertoire. LoRA allows the STG-trained backbone to remain intact for its original task, while efficiently adapting to the DA model with a minimal number of additional parameters. By leveraging LoRA, we introduce only approximately 40\% of the total number of required parameters for full retraining, reducing both storage and computational demands.

The DA model exhibits three distinct activity regimes: slow pacemaking, fast spiking (a burst that fails to repolarize), and bursting. In the following, we focus on the slow pacemaking and bursting regimes. Despite the marked differences between STG and DA neurons in terms of conductance composition, timescales, and dynamical repertoire, the adapted pipeline successfully handles both tested regimes. This demonstrates that the same organizational principles observed in the STG DIC space, namely clear separation between activity classes and the emergence of manifold structure for spiking-like regimes, generalize to a fundamentally different neuron type.

\begin{figure}[H]
  \centering
  \begin{adjustwidth}{-2.2in}{0in}
  \includegraphics[width=4.2in]{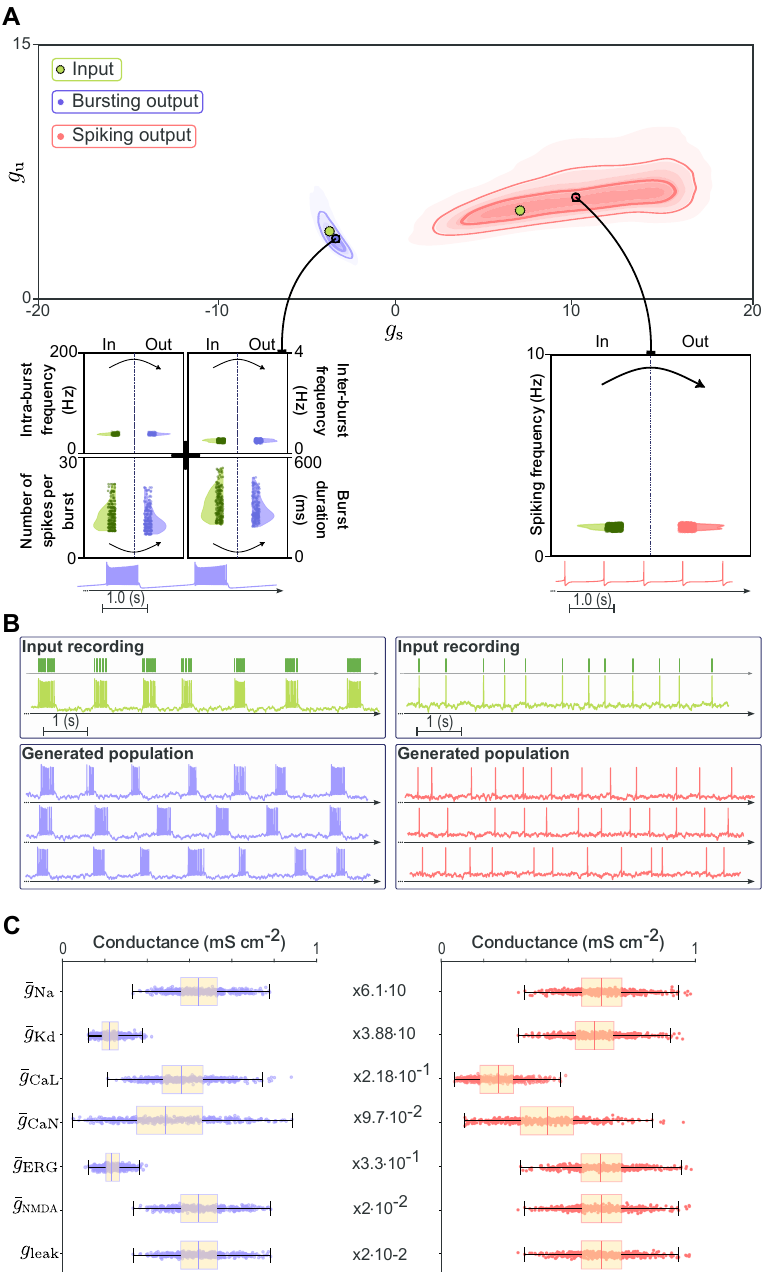}
  \caption{\textbf{Performance of the adapted pipeline on the DA neuron model.}
  \textbf{(A)} Two target DIC values (green dots) are selected in the DA $(g_\s, g_\us)$ plane, one for each tested activity regime: slow pacemaking and bursting. For each target, the pipeline infers a conditional density over DICs (contour plots), from which DIC targets are sampled to generate output populations. Violin plots compare input and output activity descriptors across regimes.
  \textbf{(B)} Target spike trains (dark green, top) compared with three representative generated spike trains for slow pacemaking (right) and bursting (left), showing accurate reproduction of activity patterns.
  \textbf{(C)} Distributions of maximal conductances across generated neurons for both regimes, demonstrating broad parameter variability despite similar dynamics.}
  \label{fig:fig_predictions_da}
  \end{adjustwidth}
\end{figure}

We evaluated the adapted pipeline using the same end-to-end protocol as for the STG model (Fig~\ref{fig:fig_predictions_da}A). Two target DIC values were selected in the DA $(g_\s, g_\us)$ plane, one per tested activity regime. For each target, an input population of $256$ neurons was generated and passed through the full pipeline to produce an output population of equal size. Across both regimes, the inferred DIC densities are well localized and the output populations closely reproduce the firing statistics of the inputs, confirming that the pipeline maintains quantitative accuracy after transfer.

Qualitative inspection further confirms that the pipeline produces spike trains consistent with the target DA activity patterns across both tested regimes (Fig~\ref{fig:fig_predictions_da}B). Representative input and generated traces show close temporal agreement for slow pacemaking and bursting. As in the STG case, the generated populations preserve degeneracy: despite producing similar output activity, the distributions of maximal conductances remain broad and heterogeneous (Fig~\ref{fig:fig_predictions_da}C), reflecting the intrinsic degeneracy of the DA conductance space.

In summary, LoRA provides an efficient strategy for extending the pipeline to new CBMs with minimal retraining. Even across neuron models with fundamentally different conductance compositions and activity repertoires, the pipeline preserves both accuracy and degeneracy, highlighting the robustness of the DIC framework as an intermediate representation for generalization across models.

\section*{Discussion}

Our results demonstrate that combining deep learning with DIC theory enables the efficient and accurate reconstruction of degenerate conductance-based populations from spike times alone. The pipeline generalizes across activity regimes, stochastic inputs mimicking physiological conditions, and distinct conductance-based models, while consistently preserving both accuracy and degeneracy. In the following, we discuss the generality of the DIC framework, its relationship to existing inference approaches, its ability to capture degeneracy, current limitations, and potential experimental applications.

\subsection*{DICs as a model-independent intermediate representation}
A key feature of our approach is that DICs provide a model-independent intermediate representation. The DIC decomposition depends only on the structure of the dynamical system, not on its particular realization, and can therefore be computed for any model admitting a separation into fast, slow, and ultra-slow feedback processes. This includes detailed conductance-based models, reduced neuronal models (e.g., Morris-Lecar or FitzHugh-Nagumo type), and analog electronic circuits implementing neuromorphic dynamics~\cite{drionDynamicInputConductances2015, mendolia2025designing}. Crucially, the first pipeline step (mapping spike times to DIC values) is entirely model-agnostic, as it operates solely on spike timing. Only the second step (generating conductance configurations from DIC constraints) requires a model-specific sensitivity matrix and compensation procedure. Extending the pipeline to a new model class therefore amounts to deriving the corresponding sensitivity matrix and implementing the appropriate compensation, while reusing the trained encoder or adapting it via transfer learning, as demonstrated with the LoRA-based adaptation from the STG to the DA model. This modularity positions the pipeline not as a tool tied to a specific model, but as a general framework applicable to any system in which neuronal excitability can be decomposed by timescale.

\subsection*{Relationship to existing inference approaches}
Our approach differs from classical SBI methods~\cite{goncalvesTrainingDeepNeural2020, cranmerFrontierSimulationbasedInference2020} and data assimilation techniques~\cite{abarbanel2009dynamical, nogaretAutomaticConstructionPredictive2016} in that it decomposes the problem through an interpretable intermediate rather than learning posteriors directly in the high-dimensional conductance space. While classical SBI methods successfully approximate posterior distributions over conductances and thereby provide families of plausible models, they do not explicitly reveal \emph{why} certain parameter combinations yield similar outputs. In contrast, the DIC representation provides a mechanistic explanation: degenerate conductance sets produce similar activity because they share similar timescale-specific feedback properties, which is precisely what DICs capture. This perspective resonates with the ``sloppy model'' framework from systems biology~\cite{gutenkunstUniversallySloppyParameter2007, transtrumPerspectiveSloppinessEmergent2015}, which shows that many parameter combinations in complex models are poorly constrained by data. In that analogy, DICs identify the neuroscience-specific ``stiff'' directions in parameter space, namely the timescale-specific feedback components that most strongly determine firing behavior, while the remaining ``sloppy'' directions correspond to the degenerate conductance combinations that the compensation algorithm explores.

\subsection*{Capturing degeneracy}
A fundamental question raised by our approach is whether the generated populations capture the ``right amount'' of degeneracy. The observed spread of maximal conductances, often varying by several-fold across instances, is consistent with biological variability reported in experimental studies of identified neurons~\cite{goaillardIonChannelDegeneracy2021, schulzVariableChannelExpression2006}. Beyond degeneracy in the conductance space, the pipeline captures degeneracy within the DIC space itself. In spiking regimes, where slow and ultra-slow components act as overlapping negative feedbacks, multiple DIC configurations produce functionally indistinguishable activity. The learned density naturally reflects this ambiguity by assigning probability mass across the corresponding one-dimensional manifold, rather than collapsing onto a single point estimate. A complete validation would require comparing inferred conductance distributions against experimental measurements from populations of neurons exhibiting similar activity, which remains an important future direction.

\subsection*{Limitations and future directions}
Despite these advantages, the method has limitations. First, its reliance on spike times makes it blind to subthreshold dynamics such as after-depolarizations or detailed interspike interval shapes. However, the DIC framework can naturally accommodate constraints at additional voltage points~\cite{drionDynamicInputConductances2015}, and building datasets that incorporate DIC values measured at multiple voltages would allow the method to infer richer neuronal dynamics. Second, the spiking regime remains inherently harder to resolve in DIC space than the bursting regime. Training under stochastic current injection exposes the network to realistic temporal variability, but does not resolve this fundamental ambiguity, since the manifold structure arises from the timescale redundancy itself rather than from insufficient input diversity. Incorporating responses to controlled stimuli ($I_\ext \ne 0$) could further constrain inference in this regime by probing the neuron under different input conditions. Third, while we currently infer only maximal conductances, this choice aligns with experimental reality where effective channel density is the primary variable under neuromodulatory or pharmacological manipulation~\cite{marder2011variability, marder2014neuromodulation}; extending the framework to kinetic parameters or structural inference remains possible in principle. Finally, at the network level, incorporating synaptic conductances into the DIC framework would account for both intrinsic and synaptic contributions to activity~\cite{prinzSimilarNetworkActivity2004}, potentially enabling real-time closed-loop neuromodulation~\cite{fyonReliableNeuromodulationAdaptive2023} or online tuning of neuromorphic devices~\cite{mendolia2025designing}.

\subsection*{Experimental and biomedical applications}
Beyond parameter inference, our pipeline offers practical tools for experimental neuroscience. Measuring all ionic conductances in a single neuron is often impractical, as typically only a subset can be probed, and the conventional approach of averaging conductance values across populations can produce fragile or unrepresentative models~\cite{golowaschFailureAveragingConstruction2002}. In contrast, our framework allows experimentalists to reconstruct plausible conductance distributions in real time from spike recordings alone. This opens concrete experimental avenues: in neuromodulation studies, comparing inferred population distributions before and after application of a neuromodulator could reveal how modulation reshapes the conductance landscape and steers neurons between activity regimes. In pharmacological settings, the pipeline could predict which ion channel targets are most affected by a given compound by tracking how inferred DIC values and conductance distributions shift under drug application. For clinical neuroscience, the ability to rapidly generate mechanistic models from extracellular recordings could support patient-specific computational modeling, for instance in the context of deep brain stimulation or epilepsy monitoring. To facilitate adoption in these diverse settings, we release the full framework as open-source software with a graphical interface~\cite{brandoit2025spike2pop}.

\subsection*{Conclusion}
In summary, by using DICs as an interpretable low-dimensional intermediate, the pipeline achieves both robustness and generalization across regimes and models. The generality of the DIC framework ensures that the approach extends to any system where excitability can be decomposed by timescale, providing a foundation for experimental strategies that probe how neurons exploit degeneracy to maintain reliable function.

\section*{Materials and methods}
\subsection*{Conductance-based models}

We use conductance-based models (CBMs) to simulate neuronal dynamics. CBMs describe the membrane potential $V$ as a function of ionic and leak currents across the membrane:
\begin{equation}
C \frac{dV}{dt} + g_\leak (V - E_\leak) = - \sum_{i \in \mathcal{I}} \bar{g}_{i} m_{i}^{p_{i}}(V, t) h_{i}^{q_{i}}(V, t) (V - E_{i}) + I_{\text{ext}}\quad.
\end{equation}
Here, $C$ is the membrane capacitance, set to $\SI{1}{\micro\farad\per\centi\meter\squared}$. Each ionic current $i \in \mathcal{I}$ is characterized by its maximal conductance $\bar{g}_{i}$, gating variables $m_{i}$ and $h_{i}$ raised to integer powers $p_{i}$ and $q_{i}$, and reversal potential $E_{i}$. The leak current $I_\leak = g_\leak (V - E_\leak)$ models passive ion flow. We denote $\bar{g} = [\bar{g}_1, \dots, \bar{g}_{|\mathcal{I}|}, g_\leak] \in \mathbb{R}_+^{N_\text{model}} = \mathcal{G}$ the vector of maximal conductances that distinguishes different instances of a model. The maximal conductances are the only parameters not treated as fixed constants in this work.

Gating variables $X \in \{m_{i}, h_{i}\}$ follow first-order voltage-dependent dynamics: 
\begin{equation} 
    \tau_X(V) \frac{dX}{dt} = X_\infty(V) - X\quad, 
\end{equation} 
where $\tau_X(V)$ is the voltage-dependent time constant and $X_\infty(V)$ is the steady-state value. To mimic physiological stochasticity, the external current $I_{\text{ext}}$ is set to a low-pass filtered Gaussian noise signal with standard deviation $\sigma_\text{noise} = \SI{5}{\micro\ampere\per\centi\meter\squared}$ and cutoff frequency $\SI{1000}{\hertz}$.

As a proof of concept, we use two established neuron models. The stomatogastric ganglion (STG) neuron model~\cite{liuModelNeuronActivityDependent1998} includes 7 ionic conductances (fast sodium, delayed rectifier potassium, calcium-dependent potassium, transient A-type potassium, slow calcium, transient T-type calcium, and hyperpolarization-activated cation) and explicitly incorporates intracellular calcium dynamics. The dopaminergic (DA) neuron model~\cite{qianMathematicalAnalysisDepolarization2014} includes 6 ionic conductances (fast sodium, delayed rectifier potassium, L-type calcium, N-type calcium, ERG potassium, and NMDA receptor-mediated). The STG model exhibits spiking and bursting, while the DA model additionally displays slow pacemaking and fast spiking. Detailed specifications and parameters for both models are provided in the Supporting Information (see \nameref{S1_appendix}).

\subsection*{Dynamic input conductances (DICs)}

DICs, introduced in~\cite{drionDynamicInputConductances2015}, provide a mathematically grounded framework for analyzing CBMs by decomposing the total membrane response into timescale-specific components.

\subsubsection*{Timescale decomposition of membrane dynamics}

The influence of membrane currents is partitioned into three characteristic temporal components. The fast dynamic conductance $g_\f(V)$ governs the rapid voltage changes underlying spike upstroke. The slow conductance $g_\s(V)$ regulates membrane repolarization and interspike interval behavior. The ultra-slow component $g_\us(V)$ accounts for adaptation and shapes the bursting envelope. Each component can be computed analytically from the CBM equations (see Supporting Information, \nameref{S1_appendix}), and the three are conveniently gathered in a sensitivity matrix $S(V)$:
\begin{equation}\label{eq:sens_matrix}
    g_\text{DICs}(V) = \begin{bmatrix} g_\f(V) \\ g_\s(V) \\ g_\us(V) \end{bmatrix} = S(V; \bar{g}) \cdot \bar{g}\quad.
\end{equation}
The sensitivity matrix summarizes the normalized influence of each conductance on the different timescales, scaled by the maximal conductances to obtain the dynamic conductances of a given instance. When $S$ is independent of $\bar{g}$, the compensatory structure is \textit{linear}; otherwise, it is \textit{nonlinear}.

\subsubsection*{Dimensionality reduction via DICs}

In the DIC framework, neuronal excitability is largely characterized by the DIC values at a critical voltage called the threshold potential $\Vth$. Following~\cite{fyonDimensionalityReductionNeuronal2024}, we approximate $\Vth$ as the first decreasing zero of the total conductance curve $g_\text{t}(V) = g_\f(V) + g_\s(V) + g_\us(V)$. This zero-crossing has a dynamical interpretation: it marks the voltage at which the net feedback from all ionic conductances changes sign, transitioning from stable (restoring) to unstable (regenerative) dynamics. In many neuron models, this is closely related to the voltage at which a saddle-node bifurcation occurs in response to sustained input~\cite{izhikevich2007dynamical}, making the neuron maximally sensitive to parameter changes and DIC values evaluated at $\Vth$ particularly informative about firing behavior. If a neuron model lacks sufficient regenerative conductances, the total conductance curve may not cross zero. In those cases, we use the mean threshold voltage across the population as a stable approximation (see Supporting Information, \nameref{S1_appendix}).

Evaluating the DICs at $\Vth$ yields a compact representation:
\begin{equation}
g_\text{DICs}(\Vth) = \begin{bmatrix} g_\f(\Vth) \\ g_\s(\Vth) \\ g_\us(\Vth) \end{bmatrix} \in \mathbb{R}^3\quad,
\end{equation}
transforming the high-dimensional parameter space $\bar{g} \in \mathcal{G}$ into a low-dimensional space. Importantly, this mapping is not injective: many distinct conductance sets can yield the same DIC vector and therefore similar activity, enabling controlled exploration of degeneracy.

\subsubsection*{Generating degenerate populations from DICs}
The method introduced in~\cite{fyonDimensionalityReductionNeuronal2024} generates maximal conductance vectors $\bar{g}$ satisfying prescribed target DIC values $g_\text{DICs}^\text{target}(\Vth)$. The conductance vector is partitioned into two subsets: $\bar{g} = [\bar{g}_\text{random}; \bar{g}_\text{comp.}]$. The random subset is sampled from distributions extending beyond the biological range~\cite{fyonDimensionalityReductionNeuronal2024, goaillardIonChannelDegeneracy2021}, while the compensable subset is determined by solving:
\begin{equation}\label{eq:linearcomp}
    S_{\text{comp.}}(\Vth) \cdot \bar{g}_{\text{comp.}} = g_{\text{DICs}}^\text{target}(\Vth) - S_{\text{random}}(\Vth) \cdot \bar{g}_{\text{random}}\quad,
\end{equation}
where the sensitivity matrix has been decomposed as $S = [S_\text{random}; S_\text{comp.}]$ following the partition of $\bar{g}$. Because different draws of $\bar{g}_\text{random}$ yield different valid solutions for $\bar{g}_\text{comp.}$, the procedure naturally produces a degenerate population from a single set of DIC constraints.

This linear compensation is exact when $S$ does not depend on $\bar{g}_\text{comp.}$. However, it becomes inaccurate for models with nonlinear compensatory structure. For example, the STG model includes intracellular calcium dynamics that depend on calcium conductances, making $S = S(V; \bar{g})$. In~\cite{fyonDimensionalityReductionNeuronal2024}, this was addressed by approximating $S$ using fixed default values, which can introduce residual errors between target and enforced DIC values. To address this, we solve the compensation iteratively: the sensitivity matrix is recomputed at each step based on the current conductance estimates, and the linear system is resolved until convergence. In practice, five iterations reduce residuals by approximately a factor of 15 compared to the single-step method, with no convergence failures observed under physiological parameter ranges. Full details of the iterative procedure, convergence analysis, and the specific conductance partitions used for each model are provided in Supporting Information (see \nameref{S1_appendix}).

\subsection*{Synthetic dataset generation} 

\begin{figure}[H] 
    \centering 
    \begin{adjustwidth}{-2.3in}{0in} 
    \includegraphics[width=7.5in]{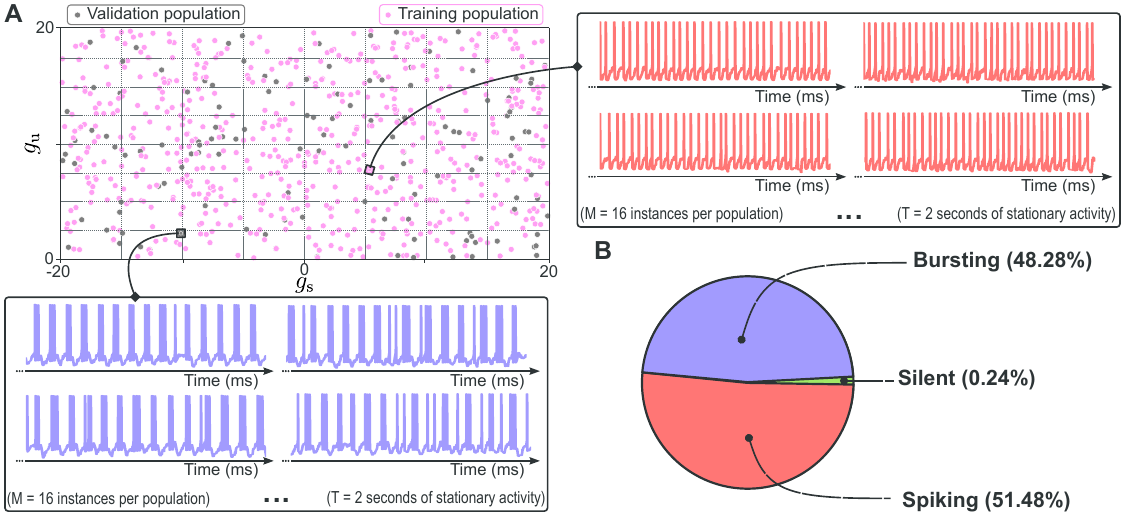} 
    \caption{{\bf The synthetic dataset generation process from sampling in the DICs space.} \textbf{(A)} Subset of the sampled DIC space used to generate degenerate CBM populations. Each dot corresponds to one population (16 instances), belonging to the training (pink) or validation (gray) set. Sampling is uniform and bounded to ensure broad coverage. Each population is simulated under noisy current injection, spike times are extracted, and descriptors are computed. Each instance is then classified as spiking, bursting, or silent. \textbf{(B)} Class distribution across the full dataset: 51.48\% spiking, 48.28\% bursting, and 0.24\% silent.} 
    \label{fig:fig_generation} 
    \end{adjustwidth} 
\end{figure}

We constructed a large open-source synthetic dataset spanning a broad range of DIC values and corresponding spike trains~\cite{brandoit2025degenerate}. Instead of sampling conductance parameters directly, we uniformly sampled the slow and ultra-slow DICs $g^* = \left(g_\s(\Vth);~g_\us(\Vth)\right)$ across bounded regions derived from extensive model exploration (Fig~\ref{fig:fig_generation}A), since these components primarily shape firing activity. For each of the $N = 75{,}000$ sampled DIC pairs, we generated a degenerate population of $M = 16$ CBM instances using the iterative compensation algorithm, yielding a total of $|\mathcal{T}| = 1{,}200{,}000$ simulated neurons. Each instance was simulated under noisy current injection, and only spike times were retained from the resulting voltage traces:
\begin{equation*} 
    V(t) \xrightarrow[]{\text{transformed into}} ~x = [t_1, t_2, \dots, t_{N_\text{spikes}}], \quad t_1 < t_2 < \dots < t_{N_\text{spikes}} \quad. 
\end{equation*} 

The final dataset consisted of 51.48\% spiking and 48.28\% bursting neurons, with 0.24\% silent instances discarded (Fig~\ref{fig:fig_generation}B). For the DA model, a reduced dataset of approximately 40\% of the STG size was used, since LoRA-based transfer learning facilitates adaptation with smaller datasets. Full details of the dataset construction, including sampling bounds, simulation parameters, activity classification criteria, and train/validation/test partitioning, are provided in Supporting Information (see \nameref{S1_appendix}).

\subsection*{Deep learning architecture and training}

\subsubsection*{Problem formulation as amortized posterior approximation}

We frame the prediction of DIC values from spike trains as an amortized posterior approximation problem~\cite{cranmerFrontierSimulationbasedInference2020}. Rather than learning a point estimate, we aim to learn a parametric conditional density $p_\theta(g^* \mid x)$ over the slow and ultra-slow DIC targets $g^* = (g_\s(\Vth);~g_\us(\Vth))$, given a variable-length spike train $x = [t_1, t_2, \dots, t_{N_\text{spikes}}]$. This formulation naturally accounts for degeneracy in the DIC space, as multiple DIC configurations may be compatible with a given spike train. At inference time, plausible DIC targets are obtained by sampling from the learned conditional density. This poses two main challenges: (1) the input sequences are of variable length, while deep neural networks typically require fixed-size inputs; and (2) learning meaningful latent representations from raw spike data, without relying on manually engineered summary statistics, is a key design objective.

To address these challenges, we design an architecture combining an attention-based encoder that maps raw spike trains into a fixed-dimensional latent representation, and a normalizing flow decoder that parameterizes the conditional density over DIC values given this representation.

\subsubsection*{Architecture overview}
The architecture comprises an attention-based encoder~\cite{vaswaniAttentionAllYou2017} and a multi-headed decoder (see Supporting Information, \nameref{S1_appendix}, for a detailed schematic and full specification).

The encoder first extracts inter-spike intervals (ISIs) and second-order ISI differences from the raw spike train, applies a logarithmic transform for numerical stability, and standardizes the resulting features. These are projected into a higher-dimensional embedding space with sinusoidal positional encoding, then processed by a stack of transformer blocks using multi-head self-attention. A self-attention pooling layer aggregates the variable-length representation into a fixed-size latent vector $z_\text{latent} \in \mathbb{R}^{d_\text{latent}}$.

The decoder operates on $z_\text{latent}$ through three parallel heads. The primary head is a RealNVP-style normalizing flow~\cite{dinh2017densityestimationusingreal} that models the conditional density $p_\theta(g^* \mid z_\text{latent})$, from which DIC targets are sampled at inference time. Two auxiliary heads, used only during training to regularize the encoder, perform classification of the firing regime (spiking or bursting) and regression of five activity descriptors (mean firing rate for spiking; intra-burst frequency, inter-burst frequency, burst duration, and spikes per burst for bursting). The auxiliary latent outputs are integrated with the encoder output via learnable element-wise mixing before being passed to the normalizing flow. At test time, only the normalizing flow head is used.

\subsubsection*{Training procedure}
The model is trained end-to-end by minimizing a composite loss combining the negative log-likelihood of the normalizing flow with weighted auxiliary losses: a masked mean squared error for the activity metrics regression and a balanced cross-entropy for firing regime classification.

To improve generalization, we apply three forms of data augmentation during training: (1) random cropping of spike trains to a window $D \sim \mathcal{U}[N_\text{spikes}/2, N_\text{spikes}]$, (2) Gaussian noise $\epsilon_i \sim \mathcal{N}(0, (2\,\text{ms})^2)$ added to spike times, and (3) 5\% spike dropout. These augmentations simulate experimental variability and are applied independently for each training sample at each update.

Hyperparameters are optimized via random search over 100 configurations~\cite{bergstraRandomSearchHyperparameter2012}, using a lightweight pointwise regression decoder in place of the normalizing flow to reduce computational cost. The selected hyperparameters are then used to train the full model with the normalizing flow head. Optimization uses AdamW~\cite{loshchilovDecoupledWeightDecay2019} and a cosine annealing learning rate schedule with warm restarts. The final architecture comprises approximately 150{,}000 trainable parameters. Full loss formulations and hyperparameter details are provided in the Supporting Information (see \nameref{S1_appendix}).

\subsubsection*{Evaluation metrics}
Model evaluation is carried out at two levels. During hyperparameter optimization, we use the MAE of the pointwise regression decoder on the validation set $\mathcal{T}_\text{val}$ as the primary metric. For test-time evaluation on $\mathcal{T}_\text{test}$, we report performance on the auxiliary tasks: the regression head is evaluated using MAE on the activity descriptors, and the classification head using balanced accuracy. These auxiliary metrics verify that the encoder learns a meaningful and structured latent representation of the input spike trains.

The quality of the predicted DIC targets is ultimately assessed through a posterior predictive check (PPC)~\cite{gelmanBayesianDataAnalysis2013}: spike trains from held-out populations are passed through the full pipeline, and the resulting generated populations are compared against the inputs using key activity descriptors. This end-to-end evaluation directly measures whether the learned conditional density produces DIC samples that, once processed by the compensation algorithm, reproduce the observed firing patterns (Fig. \ref{fig:fig_quantitative} and Fig. \ref{fig:fig_predictions_da}). We additionally assess posterior calibration in the Supporting Information (see \nameref{S1_appendix}) using TARP \cite{lemos2023samplingbasedaccuracytestingposterior}, simulation-based calibration rank histograms~\cite{taltsValidatingBayesianInference2018}, and expected coverage tests~\cite{cook2006validation}.

\subsection*{Transfer to the DA model} 
To adapt the pipeline to the DA neuron model, we apply parameter-efficient fine-tuning using Low-Rank Adaptation (LoRA)~\cite{huLoRALowRankAdaptation2021}. LoRA adapters are introduced in the linear layers of the network while attention layers remain unmodified. The majority of parameters from the STG-trained backbone are frozen, and only the newly introduced LoRA parameters are updated; the input normalization layer is recalculated based on the DA training set. The DA dataset is generated using the same procedure as for the STG model, yielding approximately 25,000 active populations after discarding silent instances. Full details are provided in Supporting Information (see \nameref{S1_appendix}).

\section*{Supporting information}

\paragraph*{S1 Appendix}
\label{S1_appendix}
\textbf{Details and additional experiments.} We provide all elements required to reproduce the results presented in this paper, as well as additional experiments. This includes model equations, DIC derivations, compensation procedure details, sampling ranges, hyperparameter tuning, architecture schematics, convergence analysis, and posterior calibration diagnostics.

\nolinenumbers

%
%

\bibliography{ref}

@article{prinzAlternativeHandTuningConductanceBased2003,
  author    = {Prinz, Astrid A. and Billimoria, Cyrus P. and Marder, Eve},
  title     = {Alternative to hand-tuning conductance-based models: construction and analysis of databases of model neurons},
  journal   = {J Neurophysiol},
  year      = {2003},
  volume    = {90},
  number    = {6},
  pages     = {3998--4015},
  doi       = {10.1152/jn.00641.2003},
}

@article{druckmannNovelMultipleObjective2007,
  author    = {Druckmann, Shaul and Banitt, Yoav and Gidon, Albert A. and Sch{\"u}rmann, Felix and Markram, Henry and Segev, Idan},
  title     = {A novel multiple objective optimization framework for constraining conductance-based neuron models by experimental data},
  journal   = {Front Neurosci},
  year      = {2007},
  volume    = {1},
  doi       = {10.3389/neuro.01.1.1.001.2007},
}

@phdthesis{shepardsonAlgorithmsInvertingHodgkinHuxley2009,
  author    = {Shepardson, Dylan},
  title     = {Algorithms for inverting {Hodgkin-Huxley} type neuron models},
  school    = {Georgia Institute of Technology},
  year      = {2009},
  type      = {Ph.D. Thesis},
  url       = {http://hdl.handle.net/1853/31686},
}

@article{huysEfficientEstimationDetailed2006,
  author    = {Huys, Quentin J. M. and Ahrens, Misha B. and Paninski, Liam},
  title     = {Efficient estimation of detailed single-neuron models},
  journal   = {J Neurophysiol},
  year      = {2006},
  volume    = {96},
  number    = {2},
  pages     = {872--890},
  doi       = {10.1152/jn.00079.2006},
}

@inproceedings{burghiDistributedOnlineEstimation2022,
  author    = {Burghi, Thiago B. and O'Leary, Timothy and Sepulchre, Rodolphe},
  title     = {Distributed online estimation of biophysical neural networks},
  booktitle = {2022 IEEE 61st Conference on Decision and Control (CDC)},
  year      = {2022},
  pages     = {628--634},
  doi       = {10.1109/CDC51059.2022.9993018},
}

@article{burghiFeedbackIdentificationConductancebased2021,
  author    = {Burghi, Thiago B. and Schoukens, Maarten and Sepulchre, Rodolphe},
  title     = {Feedback identification of conductance-based models},
  journal   = {Automatica},
  year      = {2021},
  volume    = {123},
  pages     = {109297},
  doi       = {10.1016/j.automatica.2020.109297},
}

@article{mengSequentialMonteCarlo2011,
  author    = {Meng, Liang and Kramer, Mark A. and Eden, Uri T.},
  title     = {A sequential {Monte Carlo} approach to estimate biophysical neural models from spikes},
  journal   = {J Neural Eng},
  year      = {2011},
  volume    = {8},
  number    = {6},
  pages     = {065006},
  doi       = {10.1088/1741-2560/8/6/065006},
}

@article{mengUnifiedApproachLinking2014,
  author    = {Meng, Liang and Kramer, Mark A. and Middleton, Steven J. and Whittington, Miles A. and Eden, Uri T.},
  title     = {A unified approach to linking experimental, statistical and computational analysis of spike train data},
  journal   = {PLoS One},
  year      = {2014},
  volume    = {9},
  number    = {1},
  pages     = {e85269},
  doi       = {10.1371/journal.pone.0085269},
}

@article{ditlevsenEstimationPartiallyObserved2014,
  author    = {Ditlevsen, Susanne and Samson, Adeline},
  title     = {Estimation in the partially observed stochastic {Morris--Lecar} neuronal model with particle filter and stochastic approximation methods},
  journal   = {Ann Appl Stat},
  year      = {2014},
  volume    = {8},
  number    = {2},
  pages     = {674--702},
  doi       = {10.1214/14-AOAS729},
}

@article{cranmerFrontierSimulationbasedInference2020,
  author    = {Cranmer, Kyle and Brehmer, Johann and Louppe, Gilles},
  title     = {The frontier of simulation-based inference},
  journal   = {Proc Natl Acad Sci U S A},
  year      = {2020},
  volume    = {117},
  number    = {48},
  pages     = {30055--30062},
  doi       = {10.1073/pnas.1912789117},
}

@article{goncalvesTrainingDeepNeural2020,
  author    = {Gon{\c{c}}alves, Pedro J. and Lueckmann, Jan-Matthis and Deistler, Michael and Nonnenmacher, Marcel and {\"O}cal, Kaan and Bassetto, Giacomo and Chintaluri, Chaitanya and Podlaski, William F. and Haddad, Sara A. and Vogels, Tim P. and Greenberg, David S. and Macke, Jakob H.},
  title     = {Training deep neural density estimators to identify mechanistic models of neural dynamics},
  journal   = {eLife},
  year      = {2020},
  volume    = {9},
  pages     = {e56261},
  doi       = {10.7554/eLife.56261},
}

@article{fyonDimensionalityReductionNeuronal2024,
  author    = {Fyon, Arthur and Franci, Alessio and Sacr{\'e}, Pierre and Drion, Guillaume},
  title     = {Dimensionality reduction of neuronal degeneracy reveals two interfering physiological mechanisms},
  journal   = {PNAS Nexus},
  year      = {2024},
  volume    = {3},
  number    = {10},
  pages     = {pgae415},
  doi       = {10.1093/pnasnexus/pgae415},
}

@article{liuModelNeuronActivityDependent1998,
  author    = {Liu, Zheng and Golowasch, Jorge and Marder, Eve and Abbott, L. F.},
  title     = {A model neuron with activity-dependent conductances regulated by multiple calcium sensors},
  journal   = {J Neurosci},
  year      = {1998},
  volume    = {18},
  number    = {7},
  pages     = {2309--2320},
  doi       = {10.1523/JNEUROSCI.18-07-02309.1998},
}

@article{qianMathematicalAnalysisDepolarization2014,
  author    = {Qian, Kun and Yu, Na and Tucker, Kristal R. and Levitan, Edwin S. and Canavier, Carmen C.},
  title     = {Mathematical analysis of depolarization block mediated by slow inactivation of fast sodium channels in midbrain dopamine neurons},
  journal   = {J Neurophysiol},
  year      = {2014},
  volume    = {112},
  number    = {11},
  pages     = {2779--2790},
  doi       = {10.1152/jn.00578.2014},
}

@article{drionDynamicInputConductances2015,
  author    = {Drion, Guillaume and Franci, Alessio and Dethier, Julie and Sepulchre, Rodolphe},
  title     = {Dynamic input conductances shape neuronal spiking},
  journal   = {eNeuro},
  year      = {2015},
  volume    = {2},
  number    = {1},
  doi       = {10.1523/ENEURO.0031-14.2015},
}

@article{goaillardIonChannelDegeneracy2021,
  author    = {Goaillard, Jean-Marc and Marder, Eve},
  title     = {Ion channel degeneracy, variability, and covariation in neuron and circuit resilience},
  journal   = {Annu Rev Neurosci},
  year      = {2021},
  volume    = {44},
  pages     = {335--357},
  doi       = {10.1146/annurev-neuro-092920-121538},
}

@article{marderMultipleModelsCapture2011,
  author    = {Marder, Eve and Taylor, Adam L.},
  title     = {Multiple models to capture the variability in biological neurons and networks},
  journal   = {Nat Neurosci},
  year      = {2011},
  volume    = {14},
  number    = {2},
  pages     = {133--138},
  doi       = {10.1038/nn.2735},
}

@article{goldmanGlobalStructureRobustness2001a,
  author    = {Goldman, Mark S. and Golowasch, Jorge and Marder, Eve and Abbott, L. F.},
  title     = {Global structure, robustness, and modulation of neuronal models},
  journal   = {J Neurosci},
  year      = {2001},
  volume    = {21},
  number    = {14},
  pages     = {5229--5238},
  doi       = {10.1523/JNEUROSCI.21-14-05229.2001},
}

@article{golowaschActivityDependentRegulationPotassium1999a,
  author    = {Golowasch, Jorge and Abbott, L. F. and Marder, Eve},
  title     = {Activity-dependent regulation of potassium currents in an identified neuron of the stomatogastric ganglion of the crab {Cancer borealis}},
  journal   = {J Neurosci},
  year      = {1999},
  volume    = {19},
  number    = {20},
  pages     = {RC33},
  doi       = {10.1523/JNEUROSCI.19-20-j0004.1999},
}

@article{taylorHowMultipleConductances2009,
  author    = {Taylor, Adam L. and Goaillard, Jean-Marc and Marder, Eve},
  title     = {How multiple conductances determine electrophysiological properties in a multicompartment model},
  journal   = {J Neurosci},
  year      = {2009},
  volume    = {29},
  number    = {17},
  pages     = {5573--5586},
  doi       = {10.1523/JNEUROSCI.4438-08.2009},
}

@article{yuMathematicalModelMidbrain2015,
  author    = {Yu, Na and Canavier, Carmen C.},
  title     = {A mathematical model of a midbrain dopamine neuron identifies two slow variables likely responsible for bursts evoked by {SK} channel antagonists and terminated by depolarization block},
  journal   = {J Math Neurosci},
  year      = {2015},
  volume    = {5},
  number    = {1},
  pages     = {5},
  doi       = {10.1186/s13408-015-0017-6},
}

@article{tiesingaRegulationSpikeTiming2008,
  author    = {Tiesinga, Paul and Fellous, Jean-Marc and Sejnowski, Terrence J.},
  title     = {Regulation of spike timing in visual cortical circuits},
  journal   = {Nat Rev Neurosci},
  year      = {2008},
  volume    = {9},
  number    = {2},
  pages     = {97--107},
  doi       = {10.1038/nrn2315},
}

@article{mainenReliabilitySpikeTiming1995,
  author    = {Mainen, Z. F. and Sejnowski, T. J.},
  title     = {Reliability of spike timing in neocortical neurons},
  journal   = {Science},
  year      = {1995},
  volume    = {268},
  number    = {5216},
  pages     = {1503--1506},
  doi       = {10.1126/science.7770778},
}

@article{williamsDiscoveringPreciseTemporal2020,
  author    = {Williams, Alex H. and Poole, Ben and Maheswaranathan, Niru and Dhawale, Ashesh K. and Fisher, Tucker and Wilson, Christopher D. and Brann, David H. and Trautmann, Eric M. and Ryu, Stephen and Shusterman, Roman and Rinberg, Dmitry and {\"O}lveczky, Bence P. and Shenoy, Krishna V. and Ganguli, Surya},
  title     = {Discovering precise temporal patterns in large-scale neural recordings through robust and interpretable time warping},
  journal   = {Neuron},
  year      = {2020},
  volume    = {105},
  number    = {2},
  pages     = {246--259},
  doi       = {10.1016/j.neuron.2019.10.020},
}

@article{prinzSimilarNetworkActivity2004,
  author    = {Prinz, Astrid A. and Bucher, Dirk and Marder, Eve},
  title     = {Similar network activity from disparate circuit parameters},
  journal   = {Nat Neurosci},
  year      = {2004},
  volume    = {7},
  number    = {12},
  pages     = {1345--1352},
  doi       = {10.1038/nn1352},
}

@misc{mendolia2025designing,
  author    = {Mendolia, Loris and Wen, Chenxi and Chicca, Elisabetta and Indiveri, Giacomo and Sepulchre, Rodolphe and Redout{\'e}, Jean-Michel and Franci, Alessio},
  title     = {A neuromodulable current-mode silicon neuron for robust and adaptive neuromorphic systems},
  year      = {2025},
  note      = {Preprint. Available from: https://arxiv.org/abs/2512.01133},
  doi       = {10.48550/arXiv.2512.01133},
}

@article{fyonReliableNeuromodulationAdaptive2023,
  author    = {Fyon, A. and Sacr{\'e}, P. and Franci, A. and Drion, G.},
  title     = {Reliable neuromodulation from adaptive control of ion channel expression},
  journal   = {IFAC-PapersOnLine},
  year      = {2023},
  volume    = {56},
  number    = {2},
  pages     = {458--463},
  doi       = {10.1016/j.ifacol.2023.10.1610},
}

@article{golowaschFailureAveragingConstruction2002,
  author    = {Golowasch, Jorge and Goldman, Mark S. and Abbott, L. F. and Marder, Eve},
  title     = {Failure of averaging in the construction of a conductance-based neuron model},
  journal   = {J Neurophysiol},
  year      = {2002},
  volume    = {87},
  number    = {2},
  pages     = {1129--1131},
  doi       = {10.1152/jn.00412.2001},
}

@article{marderNeuromodulationNeuronalCircuits2012,
  author    = {Marder, Eve},
  title     = {Neuromodulation of neuronal circuits: back to the future},
  journal   = {Neuron},
  year      = {2012},
  volume    = {76},
  number    = {1},
  pages     = {1--11},
  doi       = {10.1016/j.neuron.2012.09.010},
}

@misc{brandoit2025spike2pop,
  author       = {Brandoit, Julien},
  title        = {{Spike2Pop}: bridging experimental neuroscience and computational modeling},
  year         = {2025},
  note         = {Software, version 1.1.1. Available from: https://github.com/julienbrandoit/Spike2Pop---Bridging-Experimental-Neuroscience-and-Computational-Modeling},
}

@article{virtanenSciPy10Fundamental2020,
  author    = {Virtanen, Pauli and Gommers, Ralf and Oliphant, Travis E. and Haberland, Matt and Reddy, Tyler and Cournapeau, David and Burovski, Evgeni and Peterson, Pearu and Weckesser, Warren and Bright, Jonathan and {van der Walt}, St{\'e}fan J. and Brett, Matthew and Wilson, Joshua and Millman, K. Jarrod and Mayorov, Nikolay and Nelson, Andrew R. J. and Jones, Eric and Kern, Robert and Larson, Eric and Carey, C J and Polat, {\.I}lhan and Feng, Yu and Moore, Eric W. and {VanderPlas}, Jake and Laxalde, Denis and Perktold, Josef and Cimrman, Robert and Henriksen, Ian and Quintero, E. A. and Harris, Charles R. and Archibald, Anne M. and Ribeiro, Ant{\^o}nio H. and Pedregosa, Fabian and {van Mulbregt}, Paul and {SciPy 1.0 Contributors}},
  title     = {{SciPy} 1.0: fundamental algorithms for scientific computing in {Python}},
  journal   = {Nat Methods},
  year      = {2020},
  volume    = {17},
  pages     = {261--272},
  doi       = {10.1038/s41592-019-0686-2},
}

@article{shampineMATLABODESuite1997,
  author    = {Shampine, Lawrence F. and Reichelt, Mark W.},
  title     = {The {MATLAB} {ODE} suite},
  journal   = {SIAM J Sci Comput},
  year      = {1997},
  volume    = {18},
  number    = {1},
  pages     = {1--22},
  doi       = {10.1137/S1064827594276424},
}

@inproceedings{vaswaniAttentionAllYou2017,
  author    = {Vaswani, Ashish and Shazeer, Noam and Parmar, Niki and Uszkoreit, Jakob and Jones, Llion and Gomez, Aidan N. and Kaiser, {\L}ukasz and Polosukhin, Illia},
  title     = {Attention is all you need},
  booktitle = {Proceedings of the 31st International Conference on Neural Information Processing Systems},
  year      = {2017},
  pages     = {6000--6010},
  doi       = {10.48550/arXiv.1706.03762},
}

@misc{baLayerNormalization2016,
  author    = {Ba, Jimmy Lei and Kiros, Jamie Ryan and Hinton, Geoffrey E.},
  title     = {Layer normalization},
  year      = {2016},
  note      = {Preprint. Available from: https://arxiv.org/abs/1607.06450},
  doi       = {10.48550/arXiv.1607.06450},
}

@misc{hendrycksGaussianErrorLinear2023,
  author    = {Hendrycks, Dan and Gimpel, Kevin},
  title     = {{Gaussian} error linear units ({GELUs})},
  year      = {2023},
  note      = {Preprint. Available from: https://arxiv.org/abs/1606.08415},
  doi       = {10.48550/arXiv.1606.08415},
}

@article{bergstraRandomSearchHyperparameter2012,
  author    = {Bergstra, James and Bengio, Yoshua},
  title     = {Random search for hyper-parameter optimization},
  journal   = {J Mach Learn Res},
  year      = {2012},
  volume    = {13},
  pages     = {281--305},
}

@misc{loshchilovDecoupledWeightDecay2019,
  author    = {Loshchilov, Ilya and Hutter, Frank},
  title     = {Decoupled weight decay regularization},
  year      = {2019},
  note      = {Preprint. Available from: https://arxiv.org/abs/1711.05101},
  doi       = {10.48550/arXiv.1711.05101},
}

@misc{huLoRALowRankAdaptation2021,
  author    = {Hu, Edward J. and Shen, Yelong and Wallis, Phillip and Allen-Zhu, Zeyuan and Li, Yuanzhi and Wang, Shean and Wang, Lu and Chen, Weizhu},
  title     = {{LoRA}: low-rank adaptation of large language models},
  year      = {2021},
  note      = {Preprint. Available from: https://arxiv.org/abs/2106.09685},
  doi       = {10.48550/arXiv.2106.09685},
}

@misc{brandoit2025degenerate,
  author    = {Brandoit, Julien and Ernst, Damien and Drion, Guillaume and Fyon, Arthur},
  title     = {Spike-train datasets from conductance-based neuron models},
  year      = {2025},
  note      = {Dataset. Available from: Zenodo},
  doi       = {10.5281/zenodo.16912160},
}

@article{bargmannConnectomeBrainFunction2013,
  author    = {Bargmann, Cornelia I. and Marder, Eve},
  title     = {From the connectome to brain function},
  journal   = {Nat Methods},
  year      = {2013},
  volume    = {10},
  number    = {6},
  pages     = {483--490},
  doi       = {10.1038/nmeth.2451},
}

@article{mccormickNeuromodulationBrainState2020,
  author    = {McCormick, David A. and Nestvogel, Dennis B. and He, Biyu J.},
  title     = {Neuromodulation of brain state and behavior},
  journal   = {Annu Rev Neurosci},
  year      = {2020},
  volume    = {43},
  pages     = {391--415},
  doi       = {10.1146/annurev-neuro-100219-105424},
}

@misc{fyon2024neuromodulation,
  author    = {Fyon, Arthur and Drion, Guillaume},
  title     = {Neuromodulation and homeostasis: complementary mechanisms for robust neural function},
  year      = {2024},
  note      = {Preprint. Available from: https://arxiv.org/abs/2412.04172},
  doi       = {10.48550/arXiv.2412.04172},
}

@article{marder2014neuromodulation,
  author    = {Marder, Eve and O'Leary, Timothy and Shruti, Sonal},
  title     = {Neuromodulation of circuits with variable parameters: single neurons and small circuits reveal principles of state-dependent and robust neuromodulation},
  journal   = {Annu Rev Neurosci},
  year      = {2014},
  volume    = {37},
  number    = {1},
  pages     = {329--346},
  doi       = {10.1146/annurev-neuro-071013-013958},
}

@article{marder2011variability,
  author    = {Marder, Eve},
  title     = {Variability, compensation, and modulation in neurons and circuits},
  journal   = {Proc Natl Acad Sci U S A},
  year      = {2011},
  volume    = {108},
  number    = {supplement\_3},
  pages     = {15542--15548},
  doi       = {10.1073/pnas.1010674108},
}

@book{izhikevich2007dynamical,
  author    = {Izhikevich, Eugene M.},
  title     = {Dynamical systems in neuroscience},
  publisher = {MIT Press},
  year      = {2007},
  doi       = {10.7551/mitpress/2526.001.0001},
}

@inproceedings{lueckmannFlexibleStatisticalInference2017,
  author    = {Lueckmann, Jan-Matthis and Goncalves, Pedro J. and Bassetto, Giacomo and {\"O}cal, Kaan and Nonnenmacher, Marcel and Macke, Jakob H.},
  title     = {Flexible statistical inference for mechanistic models of neural dynamics},
  booktitle = {Advances in Neural Information Processing Systems},
  year      = {2017},
  volume    = {30},
  doi       = {10.1101/838383},
}

@article{nogaretAutomaticConstructionPredictive2016,
  author    = {Nogaret, Alain and Meliza, C. Daniel and Margoliash, Daniel and Abarbanel, Henry D. I.},
  title     = {Automatic construction of predictive neuron models through large scale assimilation of electrophysiological data},
  journal   = {Sci Rep},
  year      = {2016},
  volume    = {6},
  number    = {1},
  pages     = {32749},
  doi       = {10.1038/srep32749},
}

@article{gutenkunstUniversallySloppyParameter2007,
  author    = {Gutenkunst, Ryan N. and Waterfall, Joshua J. and Casey, Fergal P. and Brown, Kevin S. and Myers, Christopher R. and Sethna, James P.},
  title     = {Universally sloppy parameter sensitivities in systems biology models},
  journal   = {PLoS Comput Biol},
  year      = {2007},
  volume    = {3},
  number    = {10},
  pages     = {e189},
  doi       = {10.1371/journal.pcbi.0030189},
}

@article{transtrumPerspectiveSloppinessEmergent2015,
  author    = {Transtrum, Mark K. and Machta, Benjamin B. and Brown, Kevin S. and Daniels, Bryan C. and Myers, Christopher R. and Sethna, James P.},
  title     = {Perspective: sloppiness and emergent theories in physics, biology, and beyond},
  journal   = {J Chem Phys},
  year      = {2015},
  volume    = {143},
  number    = {1},
  pages     = {010901},
  doi       = {10.1063/1.4923066},
}

@article{schulzVariableChannelExpression2006,
  author    = {Schulz, David J. and Goaillard, Jean-Marc and Marder, Eve},
  title     = {Variable channel expression in identified single and electrically coupled neurons in different animals},
  journal   = {Nat Neurosci},
  year      = {2006},
  volume    = {9},
  number    = {3},
  pages     = {356--362},
  doi       = {10.1038/nn1639},
}

@misc{vetter2024sourcerersamplebasedmaximumentropy,
  author    = {Vetter, Julius and Moss, Guy and Schr{\"o}der, Cornelius and Gao, Richard and Macke, Jakob H.},
  title     = {{Sourcerer}: sample-based maximum entropy source distribution estimation},
  year      = {2024},
  note      = {Preprint. Available from: https://arxiv.org/abs/2402.07808},
  doi       = {10.48550/arXiv.2402.07808},
}

@article{burghiQuantitativePrediction2025,
  author    = {Burghi, Thiago B. and Schapiro, Kyra and Ivanova, Maria and Wang, Huaxinyu and Marder, Eve and O'Leary, Timothy},
  title     = {Quantitative prediction of intracellular dynamics and synaptic currents in a small neural circuit},
  journal   = {Front Comput Neurosci},
  year      = {2025},
  volume    = {19},
  doi       = {10.3389/fncom.2025.1515194},
}

@article{burghiRapidInterpretable2025,
  author    = {Burghi, Thiago B. and Ivanova, Maria and Morozova, Ekaterina and Wang, Huaxinyu and Marder, Eve and O'Leary, Timothy},
  title     = {Rapid, interpretable data-driven models of neural dynamics using recurrent mechanistic models},
  journal   = {Proc Natl Acad Sci U S A},
  year      = {2025},
  volume    = {122},
  number    = {32},
  pages     = {e2426916122},
  doi       = {10.1073/pnas.2426916122},
}

@article{fyonNewPerspective2026,
  author    = {Fyon, Arthur and Pavlova, Oleksandra and Schaar, Nick and Mesirca, Pietro and Ligot, Amandine and Gaillardon, Marvin and Brandoit, Julien and Ringlet, Sofian and Franci, Alessio and Mangoni, Matteo E. and Roeper, Jochen and Drion, Guillaume and Seutin, Vincent and Jehasse, Kevin},
  title     = {A new perspective on slow pacemaking in brain and heart},
  journal   = {bioRxiv},
  year      = {2026},
  note      = {Preprint},
  doi       = {10.1101/2025.10.30.685563},
}

@misc{hermans2022trustcrisissimulationbasedinference,
  author    = {Hermans, Joeri and Delaunoy, Arnaud and Rozet, Fran{\c{c}}ois and Wehenkel, Antoine and Begy, Volodimir and Louppe, Gilles},
  title     = {A trust crisis in simulation-based inference? {Y}our posterior approximations can be unfaithful},
  year      = {2022},
  note      = {Preprint. Available from: https://arxiv.org/abs/2110.06581},
  doi       = {10.48550/arXiv.2110.06581},
}

@misc{dinh2017densityestimationusingreal,
  author    = {Dinh, Laurent and Sohl-Dickstein, Jascha and Bengio, Samy},
  title     = {Density estimation using {Real NVP}},
  year      = {2017},
  note      = {Preprint. Available from: https://arxiv.org/abs/1605.08803},
  doi       = {10.48550/arXiv.1605.08803},
}

@book{gelmanBayesianDataAnalysis2013,
  author    = {Gelman, Andrew and Carlin, John B. and Stern, Hal S. and Dunson, David B. and Vehtari, Aki and Rubin, Donald B.},
  title     = {{Bayesian} data analysis},
  edition   = {3rd},
  publisher = {CRC Press},
  address   = {Boca Raton, FL},
  year      = {2013},
  doi       = {10.1201/9780429258411},
}

@inproceedings{lemos2023samplingbasedaccuracytestingposterior,
  author    = {Lemos, Pablo and Coogan, Adam and Hezaveh, Yashar and Perreault-Levasseur, Laurence},
  title     = {Sampling-based accuracy testing of posteriors in general-purpose inference engines},
  booktitle = {Proceedings of the 40th International Conference on Machine Learning},
  year      = {2023},
}

@misc{taltsValidatingBayesianInference2018,
  author    = {Talts, Sean and Betancourt, Michael and Simpson, Daniel and Vehtari, Aki and Gelman, Andrew},
  title     = {Validating {Bayesian} inference algorithms with simulation-based calibration},
  year      = {2018},
  note      = {Preprint. Available from: https://arxiv.org/abs/1804.06788},
  doi       = {10.48550/arXiv.1804.06788},
}

@article{cook2006validation,
  author    = {Cook, Samantha R. and Gelman, Andrew and Rubin, Donald B.},
  title     = {Validation of software for {Bayesian} models using posterior quantiles},
  journal   = {J Comput Graph Stat},
  year      = {2006},
  volume    = {15},
  number    = {3},
  pages     = {675--692},
  doi       = {10.1198/106186006X136976},
}

@article{abarbanel2009dynamical,
  author    = {Abarbanel, Henry D. I. and Creveling, Daniel R. and Farsian, Reza and Kostuk, Mark},
  title     = {Dynamical state and parameter estimation},
  journal   = {SIAM J Appl Dyn Syst},
  year      = {2009},
  volume    = {8},
  number    = {4},
  pages     = {1341--1381},
  doi       = {10.1137/090749761},
}

@article{meliza2014estimating,
  author    = {Meliza, C. Daniel and Kostuk, Mark and Huang, Hao and Nogaret, Alain and Margoliash, Daniel and Abarbanel, Henry D. I.},
  title     = {Estimating parameters and predicting membrane voltages with conductance-based neuron models},
  journal   = {Biol Cybern},
  year      = {2014},
  volume    = {108},
  number    = {4},
  pages     = {495--516},
  doi       = {10.1007/s00422-014-0615-5},
}

\end{document}


\nolinenumbers
\maketitle

We present all the elements needed in addition to the main text to reproduce the results reported in this paper. We also provide additional experiments that complement the main findings.

\section{Conductance-based models}
\label{sec:cbm_models}

This section details the equations of the conductance-based models (CBMs) used in this work. The stomatogastric ganglion (STG) model was adapted from \cite{liuModelNeuronActivityDependent1998}, and the dopaminergic (DA) model was adapted from \cite{qianMathematicalAnalysisDepolarization2014}.

Both models follow a common structure in which the membrane potential $V$ evolves according to an ordinary differential equation involving ionic and leak currents:
\begin{equation}
C \frac{dV}{dt} + g_\leak (V - E_\leak) = - \sum_{i \in \mathcal{I}} \bar{g}_{i} m_{i}^{p_{i}}(V, t) h_{i}^{q_{i}}(V, t) (V - E_{i}) + I_{\text{ext}}\quad.
\end{equation}
Here, $C$ is the membrane capacitance, set to $\SI{1}{\micro\farad\per\centi\meter\squared}$ in both models. The set $\mathcal{I}$ contains all ionic conductances considered for a given model, detailed below for each case. Each ionic current is characterized by its maximal conductance $\bar{g}_{i}$ and by gating variables $m_{i}$ (activation) and $h_{i}$ (inactivation), each raised to integer powers $p_{i}$ and $q_{i}$, respectively. $E_{i}$ is the Nernst reversal potential for the associated ion. The leak current is modeled as $I_\leak = g_\leak (V - E_\leak)$. The external input current $I_{\text{ext}}$ is set to a low-pass filtered Gaussian noise signal with standard deviation $\sigma_\text{noise} = \SI{5}{\micro\ampere\per\centi\meter\squared}$ and cutoff frequency $\SI{1000}{\hertz}$, as described in the main text. We denote $\bar{g} = [\bar{g}_1,~ \bar{g}_2,~ \dots,~ \bar{g}_{|\mathcal{I}|}, g_\leak] \in \mathbb{R}^{N_\text{model}}$ the vector of maximal conductances (extended by the leak conductance) that defines different instances of a model. The maximal conductances are the only parameters varied in this work.

Gating variables $X \in \{m_{i}, h_{i}\}$ are dimensionless quantities constrained between 0 and 1, representing the fraction of ion channels that are activated ($m_i$) or not inactivated ($h_i$). They follow first-order voltage-dependent dynamics of the form: 
\begin{equation}
\tau_X(V) \frac{dX}{dt} = X_\infty(V) - X\quad,
\end{equation}
where $\tau_X(V)$ is the voltage-dependent time constant and $X_\infty(V)$ is the steady-state activation or inactivation value.

Simulations are implemented in Python using the SciPy library~\cite{virtanenSciPy10Fundamental2020}, solving the system of differential equations with the BDF solver~\cite{shampineMATLABODESuite1997}, well suited for stiff systems with multiple timescales. The maximum time step is set to $\SI{0.05}{\milli\second}$. Simulations are parallelized across CPU cores to accelerate dataset generation.

\subsection{The stomatogastric ganglion neuron model} \label{appendix:stg_model_eqs}
The STG model~\cite{liuModelNeuronActivityDependent1998} includes eight ionic currents: the fast transient sodium current ($\Na$), the delayed rectifier potassium current ($\Kd$), the calcium-activated potassium current ($\KCa$), the A-type potassium current ($\A$), the slow calcium current ($\CaS$), the transient calcium current ($\CaT$), the hyperpolarization-activated current ($\Hh$), and the leak current. In addition to the standard voltage-dependent gating, this model explicitly incorporates intracellular calcium concentration dynamics through an additional ordinary differential equation:
$$
\tau_\Ca \dfrac{d\Ca}{dt} = - \alpha_\Ca \left(I_\CaS + I_\CaT\right) - \Ca + \beta_\Ca\quad,
$$
which modulates calcium-dependent potassium currents through a voltage- and calcium-dependent steady-state gating function $m_{\infty, \KCa}(V, \Ca)$. This coupling introduces a nonlinear dependence between conductance parameters and the sensitivity matrix, motivating the iterative compensation algorithm described in Section~\ref{sec:pop_generation}.

Table~\ref{table:appendix_stg_parameters} provides the values of the fixed model parameters. As initial conditions, we used $V_0 = \SI{-70}{\milli\volt}$ and $\Ca_0 = \SI{0.5}{\uM}$, with gating variables initialized at steady state: $X_0 = X_\infty(V_0; \Ca_0)$. The first $\SI{3000}{\milli\second}$ of each simulation were discarded to avoid transient effects. Table~\ref{table:appendix_stg_eq} lists the mathematical expressions for the steady-state gating functions, the time constant functions, and the corresponding gating exponents.

\begin{table}[!htb]
\caption[Fixed parameters for the STG model.]{\textbf{Fixed parameters for the STG model.} Reversal potentials, calcium time constant, and calcium dynamics parameters used in the STG model.}
    \label{table:appendix_stg_parameters}
    \centering
    \begin{tabular}{cccccccc}
        \toprule
        $E_{\text{leak}}$ & $E_{\Na}$ & $E_{\text{K}}$ & $E_{\Hh}$ & $E_{\Ca}$ & $\tau_{\text{Ca}}$ & $\alpha_{\text{Ca}}$ & $\beta_\Ca$\\
        \midrule
        $\SI{-50}{\milli\volt}$ & $\SI{50}{\milli\volt}$ & $\SI{-80}{\milli\volt}$ & $\SI{-20}{\milli\volt}$ & $\SI{80}{\milli\volt}$ & $\SI{20}{\milli\second}$ & $\SI{0.94}{\mM \nano\farad \per \nano \ampere}$ & $\SI{0.05}{\uM}$\\
        \bottomrule
    \end{tabular}
\end{table}

\subsection{The dopaminergic neuron model}  \label{appendix:da_model_eqs}
The DA model~\cite{qianMathematicalAnalysisDepolarization2014} includes seven ionic currents: the fast sodium current ($\Na$), the delayed rectifier potassium current ($\Kd$), the ERG potassium current ($\ERG$), the L-type calcium current ($\CaL$), the N-type calcium current ($\CaN$), the NMDA receptor-mediated current ($\NMDA$), and the leak current. This model incorporates a magnesium-sensitive current through the NMDA channel, influenced by a fixed extracellular magnesium concentration.

The NMDA current is treated as instantaneous and always evaluated at its steady-state value:
$$
I_{\text{NMDA}} = \bar{g}_{\text{NMDA}} (V - E_{\text{NMDA}}) \cdot m_{\text{NMDA},\infty}(V, \Mg)\quad,
$$
where $m_{\text{NMDA},\infty}$ is the voltage- and magnesium-dependent steady-state activation function.

For the ERG channel, the gating variables $o_{\text{ERG}}$ and $i_{\text{ERG}}$ evolve according to the differential equations:
$$
\frac{d o_{\text{ERG}}}{dt} = a_0(V) \cdot (1 - o_{\text{ERG}} - i_{\text{ERG}}) + b_i(V) \cdot i_{\text{ERG}} - o_{\text{ERG}} \cdot (a_i(V) + b_0(V))\quad,
$$
$$
\frac{d i_{\text{ERG}}}{dt} = a_i(V) \cdot o_{\text{ERG}} - b_i(V) \cdot i_{\text{ERG}}\quad.
$$
The steady-state expressions for the ERG gating variables under constant voltage $V$ are:
$$
o_{\text{ERG},\infty}(V) = \frac{a_0(V) \cdot b_i(V)}{a_0(V)\cdot(a_i(V) + b_i(V)) + b_0(V) \cdot b_i(V)}\quad,
$$
$$
i_{\text{ERG},\infty}(V) = \frac{a_0(V) \cdot a_i(V)}{a_0(V)(a_i(V) + b_i(V)) + b_0(V) \cdot b_i(V)}\quad.
$$
The corresponding ERG current is:
$$
I_{\text{ERG}} = \bar{g}_{\text{ERG}} \cdot o_{\text{ERG}} \cdot (V - E_K)\quad.
$$

Table~\ref{table:appendix_da_parameters} provides the fixed model parameters. As initial conditions, we used $V_0 = \SI{-90}{\milli\volt}$, with gating variables initialized at steady state: $X_0 = X_\infty(V_0)$. The first $\SI{3000}{\milli\second}$ of each simulation were discarded to avoid transient effects. Table~\ref{table:appendix_da_eq} lists the mathematical expressions for all gating functions and kinetic parameters.

\begin{table}[!htb]
    \caption[Fixed parameters for the DA model.]{\textbf{Fixed parameters for the DA model.} Reversal potentials and magnesium concentration used in the DA model.}
    \label{table:appendix_da_parameters}
    \centering
    \begin{tabular}{ccccccc}
        \toprule
        $E_{\Na}$ & $E_{\text{K}}$ & $E_{\Ca}$ & $E_{\text{leak}}$ & $E_{\text{NMDA}}$ & Mg \\
        \midrule
        \SI{60}{\milli\volt} & \SI{-85}{\milli\volt} & \SI{60}{\milli\volt} & \SI{-50}{\milli\volt} & \SI{0}{\milli\volt} & \SI{1.4}{} \\
        \bottomrule
    \end{tabular}
\end{table}

\begin{landscape}

\begin{table}[!htb]
    \caption[Gating functions and kinetics for the STG model.]{\textbf{Gating functions and kinetics for the STG model.} Steady-state activation/inactivation functions, time constants, and exponents for each ionic current in the STG model. All $f$ functions refer to the generalized sigmoid function (Eq.~\ref{eq:gsig}).}
    \label{table:appendix_stg_eq}
    \centering
    
    \begin{tabular}{lcccccc}
        \toprule
        Current $I_i$ & $p_i$ &$q_i$& $m_{i, \infty}(V)$ or $m_{i, \infty}(V, \Ca)$ & $h_{i, \infty}(V)$ & $\tau_{m_i}(V)$ & $\tau_{h_i}(V)$ \\
        \midrule
        $I_{\Na}$  & 3 & 1 &
        $\mathord{\scriptstyle f(V, 0, 1, -5.29, 25.5)}$ &
        $\mathord{\scriptstyle f(V, 0, 1, 5.18, 48.9)}$ &
        $\mathord{\scriptstyle f(V, 1.32, -1.26, -25, 120)}$ &
        $\mathord{\scriptstyle f(V, 0, 0.67, -10, 62.9)} \cdot \mathord{\scriptstyle f(V, 1.5, 1, 3.6, 34.9)}$ \\

        $I_{\Kd}$  & 4 & 0 &
        $\mathord{\scriptstyle f(V, 0, 1, -11.8, 12.3)}$ & --- &
        $\mathord{\scriptstyle f(V, 7.2, -6.4, -19.2, 28.3)}$ & --- \\

        $I_{\CaT}$  & 3 & 1 &
        $\mathord{\scriptstyle f(V, 0, 1, -7.2, 27.1)}$ &
        $\mathord{\scriptstyle f(V, 0, 1, 5.5, 32.1)}$ &
        $\mathord{\scriptstyle f(V, 21.7, -21.3, -20.5, 68.1)}$ &
        $\mathord{\scriptstyle f(V, 105, -89.8, -16.9, 55)}$ \\

        $I_{\CaS}$  & 3 & 1 &
        $\mathord{\scriptstyle f(V, 0, 1, -8.1, 33)}$ &
        $\mathord{\scriptstyle f(V, 0, 1, 6.2, 60)}$ &
        $ \mathord{\scriptstyle1.4 + \frac{7}{\exp\left(\frac{V+27}{10}\right) + \exp\left(\frac{V+70}{-13}\right)}}$ &
        $ \mathord{\scriptstyle60 + \frac{150}{\exp\left(\frac{V+55}{9}\right) + \exp\left(\frac{V+65}{-16}\right)}}$ \\

        $I_{\KCa}$  & 4 & 0 &
        $\mathord{\scriptstyle\frac{\text{Ca}}{\text{Ca} + 3} \cdot f(V, 0, 1, -12.6, 28.3)}$ & --- &
        $\mathord{\scriptstyle f(V, 90.3, -75.1, -22.7, 46)}$ & --- \\

        $I_\A$  & 3 & 1 &
        $\mathord{\scriptstyle f(V, 0, 1, -8.7, 27.2)}$ &
        $\mathord{\scriptstyle f(V, 0, 1, 4.9, 56.9)}$ &
        $\mathord{\scriptstyle f(V, 11.6, -10.4, -15.2, 32.9)}$ &
        $\mathord{\scriptstyle f(V, 38.6, -29.2, -26.5, 38.9)}$ \\

        $I_\Hh$  & 1 & 0 &
        $\mathord{\scriptstyle f(V, 0, 1, 6, 70)}$ & --- &
        $\mathord{\scriptstyle f(V, 272, 1499, -8.73, 42.2)}$ & --- \\

        $I_\leak$  & 0 & 0 & --- & --- & --- & --- \\
        \bottomrule
    \end{tabular}

    \vspace{1em}

    \caption[Gating functions and kinetics for the DA model.]{
\textbf{Gating functions and kinetics for the DA model.} 
All $f$ functions refer to the generalized sigmoid function (Eq~\ref{eq:gsig}). 
Additionally, there is a current $I_\mathrm{ERG}$ described by the ERG channel, whose rate functions are defined as exponentials of the membrane potential $V$. 
The activation and inactivation parameters are as follows: 
$a_0(V) = 0.0036 \exp(0.0759V)$, 
$b_0(V) = 1.2523 \times 10^{-5} \exp(-0.0671V)$, 
$a_i(V) = 0.1 \exp(0.1189V)$, 
and $b_i(V) = 0.003 \exp(-0.0733V)$.
}
\label{table:appendix_da_eq}

    \centering
    \setlength{\tabcolsep}{4pt} 
    \begin{tabular}{lcccccc}
        \toprule
        Current & $p_i$ & $q_i$ & $m_{i, \infty}(V)$ or $m_{i, \infty}(V, \Mg)$ & $h_{i, \infty}(V)$ & $\tau_{m_i}(V)$ & $\tau_{h_i}(V)$ \\
        \midrule
        $I_{\Na}$  & 3 & 1 &
        $\mathord{\scriptstyle f(V, 0, 1, -9.7264, 30.0907)}$ &
        $\mathord{\scriptstyle f(V, 0, 1, 10.7665, 54.0289)}$ &
        $\mathord{\scriptstyle 0.01 + \frac{1.0}{(-\frac{15.6504 + 0.4043V}{\exp(-19.565 - 0.5052V) - 1.0}) + 3.0212 \exp(-0.007463V)}}$ &
        $\mathord{\scriptstyle 0.4 + \frac{1.0}{(0.00050754 \exp(-0.063213V)) + 9.7529 \exp(0.13442V)}}$ \\

        $I_{\Kd}$  & 3 & 0 &
        $\mathord{\scriptstyle f(V, 0, 1, -12, 25)}$ & --- &
        $\mathord{\scriptstyle f(V, 20, -18, -10, 38)}$ & --- \\

        $I_{\CaL}$  & 2 & 0 &
        $\mathord{\scriptstyle f(V, 0, 1, -2, 50)}$ & --- &
        $\mathord{\scriptstyle f(V, 30, -28, -3, 45)}$ & --- \\

        $I_{\CaN}$  & 1 & 0 &
        $\mathord{\scriptstyle f(V, 0, 1, -7, 30)}$ & --- &
        $\mathord{\scriptstyle f(V, 30, -25, -6, 55)}$ & --- \\

        $I_{\NMDA}$  & 1 & 0 &
        $\mathord{\scriptstyle \frac{1}{1 + \frac{\Mg \cdot \exp(-0.08V)}{10}}}$ & --- &
        --- & --- \\
        
        $I_{\leak}$  & 0 & 0 & --- & --- & --- & --- \\

        \bottomrule
    \end{tabular}

\end{table}

\begin{equation}\label{eq:gsig}
    f(V, A, B, C, D) = A + \frac{B}{1 + \exp\left(\frac{V + D}{C}\right)}
\end{equation}

\end{landscape}

\section{Dynamic input conductances (DICs)}
\label{sec:dics}

This work uses the concept of Dynamic Input Conductances (DICs)~\cite{drionDynamicInputConductances2015}, which consist of voltage-dependent conductances separated according to timescales. We use three timescales: fast, slow, and ultra-slow. These three DIC components have been shown to be sufficient to qualitatively determine excitability in neuron models~\cite{drionDynamicInputConductances2015, fyonDimensionalityReductionNeuronal2024}. Specifically, based on DIC values at threshold, it becomes possible to predict the firing pattern of a neuron. The computation follows recent work on the subject~\cite{fyonReliableNeuromodulationAdaptive2023, fyonDimensionalityReductionNeuronal2024}.

We denote $g_\f(V)$, $g_\s(V)$, and $g_\us(V)$ the fast, slow, and ultra-slow DIC components, respectively. The total DIC is $g_\text{t}(V) = g_\f(V) + g_\s(V) + g_\us(V)$. In this work, we focus on a particular voltage value, the threshold voltage $\Vth$. As an approximation, we consider $\Vth$ to be a fixed value shared by all instances of a given CBM. We make this approximation mainly because computing $\Vth$ requires knowing $\bar{g}$, which is unknown in advance. The choice of $\Vth$ is discussed in Section~\ref{sec:vth_approx}.

\subsection{Computation details}
In CBMs, the timescale-specific conductances can be computed analytically:
\begin{equation}\label{eq:dics_cbm}
\begin{cases}
    g_\f(V)  &= \left.\left[-\dfrac{\partial \dot{V}}{\partial V}- \sum_i w_{\f\s, X_i}(V) \left(\dfrac{\partial \dot{V}}{\partial X_i}\dfrac{\partial X_{i,\infty}}{\partial V}\right)\right]\right|_{V}\dfrac{1}{g_\leak}\quad,\\[8pt]
    g_\s(V)  &= \left.\left[- \sum_i \left(w_{\s\us, X_i}(V) - w_{\f\s, X_i}(V)\right) \left(\dfrac{\partial \dot{V}}{\partial X_i}\dfrac{\partial X_{i,\infty}}{\partial V}\right)\right]\right|_{V}\dfrac{1}{g_\leak}\quad,\\[8pt]
    g_\us(V) &= \left.\left[- \sum_i \left(1 - w_{\s\us, X_i}(V)\right) \left(\dfrac{\partial \dot{V}}{\partial X_i}\dfrac{\partial X_{i,\infty}}{\partial V}\right)\right]\right|_{V}\dfrac{1}{g_\leak}\quad,
    \end{cases}
\end{equation}
where the terms $X_i$ correspond to gating variables ($m_i$ or $h_i$) controlling activation and inactivation of ion channels, and $w_{\f\s, X_i}(V)$ and $w_{\s\us, X_i}(V)$ are voltage-dependent weighting factors. The voltage dependence is omitted in the right-hand sides for readability. The sign convention and normalization by the leak conductance follow~\cite{fyonDimensionalityReductionNeuronal2024}, and thus differ from the original formulation in~\cite{drionDynamicInputConductances2015}.

\subsubsection*{Weighting functions}
The weighting functions determine how each gating variable contributes to different timescales, based on a logarithmic scaling between 0 and 1 using reference timescales:
\begin{align}\label{eq:w_factors_dics}
    w_{\f\s, X_i}(V) &= \begin{cases}
        1&,\quad \tau_{X_i}(V) \le \tau_\f(V)\quad ,\\
        \dfrac{\log\left(\tau_\s(V)\right) - \log\left(\tau_{X_i}(V)\right)}{\log\left(\tau_\s(V)\right) - \log\left(\tau_\f(V)\right)}&,\quad \tau_\f(V) < \tau_{X_i}(V) \le \tau_\s(V) \quad ,\\
        0&,\quad \tau_{X_i}(V) > \tau_\s(V)\quad ,
    \end{cases}\notag\\
\vspace{12pt}\\
    w_{\s\us, X_i}(V) &= \begin{cases}
        1&,\quad \tau_{X_i}(V) \le \tau_\s(V)\quad ,\\
        \dfrac{\log\left(\tau_\us(V)\right) - \log\left(\tau_{X_i}(V)\right)}{\log\left(\tau_\us(V)\right) - \log\left(\tau_\s(V)\right)}&,\quad \tau_\s(V) < \tau_{X_i}(V) \le \tau_\us(V)\quad , \\
        0&,\quad \tau_{X_i}(V) > \tau_\us(V)\quad.
    \end{cases}\notag
\end{align}
The reference timescales $\tau_\f$, $\tau_\s$, and $\tau_\us$ are chosen according to the characteristic timescales of bursting and spiking dynamics~\cite{drionDynamicInputConductances2015}. Specifically, $\tau_\f$ corresponds to the activation time constant of the fastest depolarizing current, $\tau_\s$ represents that of the fastest repolarizing current, and $\tau_\us$ is associated with the slowest variable in the system, typically governing burst adaptation.

For the STG model, we used $\tau_{m_\Na}(V)$, $\tau_{m_\Kd}(V)$, and $\tau_{\Hh}(V)$, respectively. For the DA model, we used $\tau_{m_\Na}(V)$, $\tau_{m_\Kd}(V)$, and a constant function $\tau_{\us, \text{DA}} = \SI{100}{\milli\second}$.

\subsubsection*{Partial derivatives for standard channels}
In the standard case of purely voltage-dependent ion channels, the terms $\frac{\partial \dot{V}}{\partial X_i}$ take the form:
\begin{equation}\label{eq:dVdX}
    \dfrac{\partial \dot{V}}{\partial m_i} = \bar{g}_i p_i m_{i, \infty}^{p_i - 1}h^{q_i}_{i, \infty} \quad;\qquad \dfrac{\partial \dot{V}}{\partial h_i} = \bar{g}_i q_i m_{i, \infty}^{p_i}h^{q_i - 1}_{i, \infty}\quad.
\end{equation}

\subsubsection*{Sensitivity matrix formulation}
A compact way of writing Eq.~\ref{eq:dics_cbm} is through the sensitivity matrix $S(V;\bar{g})$:
\begin{equation}\label{eq:sens_matrix_si}
    g_\text{DICs}(V) = \begin{bmatrix}
        g_\f(V)\\g_\s(V)\\g_\us(V)
    \end{bmatrix} = S(V;\bar{g})\cdot\bar{g}\quad.
\end{equation}
This is the formulation used in the generation procedure. One can construct the sensitivity matrix element by element by computing:
\begin{equation}
 S_{j,i} = \dfrac{g^{(i)}_j}{\bar{g}_i},\quad (i, j) \in \{1, 2, \dots, N_{\text{model}}\}\times\{\f, \s, \us\}\quad,
\end{equation}
where $g^{(i)}_j$ corresponds to the partial sum from Eq.~\ref{eq:dics_cbm} involving only current $i$ (that is, keeping only the terms containing $\bar{g}_{i}$, as in Eq.~\ref{eq:dVdX}) for the DIC component associated with timescale $j$.

When $S$ is independent of $\bar{g}$, the compensatory structure is \textit{linear}; otherwise, it is \textit{nonlinear}. In the STG model, intracellular calcium dynamics introduce a dependence of $S$ on calcium conductances, resulting in nonlinear compensatory structure. The DA model has linear compensatory structure.

\section{Population generation procedure}
\label{sec:pop_generation}

In this work, we improved the procedure described in~\cite{fyonDimensionalityReductionNeuronal2024} to generate degenerate populations of CBMs. The improved method, called the iterative compensation algorithm, is detailed below. The procedure is used both during dataset generation (Fig~6 from the main text) and at inference time once the deep learning architecture has been trained (Fig~3--6 from the main text). The core idea is to impose DIC values at the threshold voltage, as these are known to shape the firing pattern~\cite{fyonDimensionalityReductionNeuronal2024, drionDynamicInputConductances2015}.

\subsection{Compensation framework}

The compensation procedure exploits the relationship $g_\text{DICs}(\Vth) = S(\Vth; \bar{g}) \cdot \bar{g}$ between DICs and maximal conductances. The conductance vector is partitioned into two subsets: $\bar{g} = [\bar{g}_\text{random};~ \bar{g}_\text{comp.}]$, where this partition preserves the total number of conductances, i.e., both subsets together contain all $N_\text{model}$ components of $\bar{g}$. The random subset $\bar{g}_\text{random}$ is sampled from distributions extending beyond the biological range~\cite{fyonDimensionalityReductionNeuronal2024, goaillardIonChannelDegeneracy2021}, while the compensable subset $\bar{g}_\text{comp.}$ is adjusted to satisfy the DIC constraints by solving:
\begin{equation}\label{eq:linearcomp_si}
    S_{\text{comp.}}(\Vth) \cdot \bar{g}_{\text{comp.}} = g_{\text{DICs}}^\text{target}(\Vth) - S_{\text{random}}(\Vth) \cdot \bar{g}_{\text{random}}\quad,
\end{equation}
where the sensitivity matrix has been decomposed as $S = [S_\text{random};~ S_\text{comp.}]$ following the partition of $\bar{g}$. Because different draws of $\bar{g}_\text{random}$ yield different valid solutions for $\bar{g}_\text{comp.}$, the procedure naturally produces a degenerate population from a single set of DIC constraints.

This linear compensation is exact when $S$ does not depend on $\bar{g}_\text{comp.}$. However, it becomes inaccurate for models with nonlinear compensatory structure. In the STG model, intracellular calcium dynamics depend on calcium conductances, making $S = S(V; \bar{g})$. In~\cite{fyonDimensionalityReductionNeuronal2024}, this was addressed by approximating $S$ using fixed default values, which can introduce residual errors between target and enforced DIC values.

\subsubsection*{Iterative extension}

To improve constraint satisfaction, we solve the compensation iteratively: starting from an initial guess $\bar{g}^{(0)}_\text{comp.}$, the sensitivity matrix is recomputed at each iteration based on the current conductance estimates, and a new linear system is solved:
\begin{equation}\label{eq:itercomp_si}
    A(\bar{g}^{(k)}_\text{comp.})\cdot \bar{g}^{(k+1)}_\text{comp.} = b(\bar{g}^{(k)}_\text{comp.}), \quad k=0,\dots,K-1\quad,
\end{equation}
where $A := S_{\text{comp.}}(\Vth)$ and $b := g_{\text{DICs}}^\text{target}(\Vth) - S_{\text{random}}(\Vth) \cdot \bar{g}_{\text{random}}$ are updated at each iteration. The process is repeated until the residual norm between target and actual DIC values becomes sufficiently small.

We chose this iterative fixed-point approach over general-purpose nonlinear optimization methods for several reasons. First, under physiological conditions, the procedure reliably converges within a small number of iterations ($K=5$ in this work), making more sophisticated techniques unnecessary. Second, each iteration requires only solving a linear system, which is computationally inexpensive. Third, the method naturally preserves the structure of the original compensation framework, ensuring that the generated populations remain consistent with DIC theory. In extensive testing across both the STG and DA models, we observed no convergence failures under physiological parameter ranges.

\subsubsection*{Two-step compensation structure}

The full generation procedure consists of two compensation steps performed in series. In the first step, $n=3$ DIC constraints (fast, slow, and ultra-slow) are imposed to ensure spontaneous activity~\cite{fyonDimensionalityReductionNeuronal2024}. For the STG model, this step adjusts one conductance contributing predominantly to each timescale: $\bar{g}_\text{comp., spont.} = (g_\Na, g_\Kd, g_\Hh)$, corresponding to fast, slow, and ultra-slow feedback respectively, while the remaining conductances are randomly sampled. For the DA model, the compensated set is $\bar{g}_\text{comp., spont.} = (g_\Na, g_\CaN, g_\ERG)$.

The second step refines the slow and ultra-slow DIC values toward the target activity ($n=2$ constraints). Starting from the conductance vector obtained after the first step, a new partition $\bar{g} = [\bar{g}_\text{random}';~ \bar{g}_\text{comp.}']$ is defined. Crucially, this partition can differ from the first step: conductances that were compensated initially may now be held fixed (becoming part of $\bar{g}_\text{random}'$), while different conductances are adjusted to satisfy the new constraints. This flexibility allows the method to target specific DIC components independently. In the STG model, the second step adjusts $\bar{g}_\text{comp.}' = (g_\CaS, g_\Hh)$ when $g_\s < 0$ (bursting regime) or $\bar{g}_\text{comp.}' = (g_\A, g_\Hh)$ when $g_\s > 0$ (spiking regime). In the DA model, the second step adjusts $\bar{g}_\text{comp.}' = (g_\ERG, g_\CaL)$ when $g_\s < 0$ or $\bar{g}_\text{comp.}' = (g_\ERG, g_\Kd)$ when $g_\s > 0$. Since this step starts from the output of the first compensation, no additional random sampling is performed.

\subsection{Validation of the iterative compensation algorithm}

To validate the iterative compensation algorithm, we compared its performance against the single-step linear compensation method from~\cite{fyonDimensionalityReductionNeuronal2024}. We evaluated both methods on 1,633 target DIC configurations, generating $P = 250$ instances per target. For each target, we computed the mean residual norm across all generated instances, defined as the Euclidean distance between the target DICs and those actually enforced by the generated conductance vector:
\begin{equation}\label{eq:residual_norm}
\|r\|_2 = \left\|g_{\text{DICs}}^\text{target}(\Vth) - S(\Vth; \bar{g}) \cdot \bar{g}\right\|_2\quad.
\end{equation}

\begin{figure}[H]
  \centering
  \includegraphics[width=0.8\textwidth]{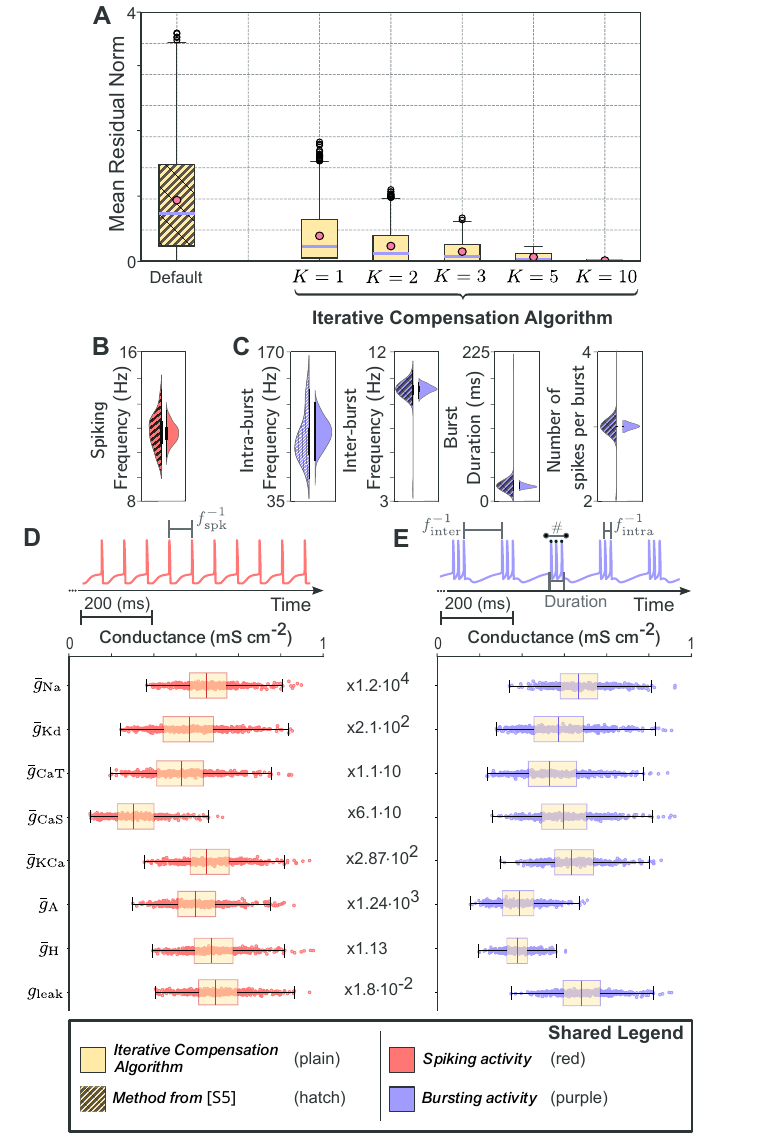}
  \caption{
\textbf{Iterative compensation improves constraint satisfaction and preserves degeneracy in CBMs with nonlinear dynamics.} 
\textbf{(A)} Mean residuals of DIC constraints across 1,633 targets and 250 instances per population, comparing linear compensation (hatched, 0 iterations from \cite{fyonDimensionalityReductionNeuronal2024}) to iterative compensation (plain), showing a rapid reduction in residuals. 
\textbf{(B)} Distribution of mean firing frequency $f_\text{spk}$ in spiking neurons (red), demonstrating tighter and more consistent activity with iterative compensation (plain) compared to linear (hatched). 
\textbf{(C)} Distributions of bursting features (intra-burst frequency $f_\text{intra}$, inter-burst frequency $f_\text{inter}$, burst duration, spikes per burst $\#$) in bursting neurons (purple), showing more compact activity profiles with iterative compensation. 
\textbf{(D, E)} Maximal conductance values across spiking (D) and bursting (E) neuron populations generated with iterative compensation, revealing high variability despite similar firing patterns, illustrating preserved degeneracy.
}
  \label{fig:fig_ICA_SI}
\end{figure}

Fig~\ref{fig:fig_ICA_SI}A compares the mean residual norms obtained with the linear method (hatched, corresponding to 0 iterations) and the iterative compensation algorithm (plain) over up to 10 iterations. After only five iterations, residuals were reduced by a factor of 15, striking a balance between accuracy and computational cost. Accordingly, we used five iterations as the default throughout the study.

Since DICs serve as an intermediate representation rather than the final target, a more direct measure of accuracy is given by the firing pattern distributions of the generated populations. Fig~\ref{fig:fig_ICA_SI}B--C compare populations generated with the linear (hatched) and iterative (plain) methods for spiking neurons (red, Fig~\ref{fig:fig_ICA_SI}B) and bursting neurons (purple, Fig~\ref{fig:fig_ICA_SI}C). The iterative algorithm reduces outliers and produces more compact, centered distributions of activity statistics. Spiking activity is summarized by mean firing frequency, while bursting activity is described by intra-burst frequency, inter-burst frequency, burst duration, and spikes per burst. These tighter distributions confirm that the iterative approach enhances the reliability of the DIC-based generative method.

Importantly, the iterative compensation algorithm preserves degeneracy. Sampling of $\bar{g}_\text{random}$ introduces variability across instances, resulting in distinct compensation problems even for the same target DIC. This ensures that maximal conductance vectors remain heterogeneous, often differing by several folds, while still producing consistent activity statistics. This is illustrated in both spiking (Fig~\ref{fig:fig_ICA_SI}D) and bursting (Fig~\ref{fig:fig_ICA_SI}E) populations, where conductance distributions span wide ranges despite similar firing patterns.

These results demonstrate that the iterative compensation algorithm substantially improves constraint satisfaction while preserving the intrinsic biological heterogeneity characteristic of degenerate neuronal populations. The method forms the basis for generating the synthetic training dataset and is also employed during inference to map predicted DIC values to conductance-based model populations.

\subsection{Approximation of the threshold voltage}
\label{sec:vth_approx}

The generation procedure requires evaluating sensitivities at a fixed voltage $\Vth$. However, the actual threshold voltage for a given conductance vector $\bar{g}$ is unknown before $\bar{g}$ is determined. We therefore approximate $\Vth$ by a constant value shared across all instances of a given CBM.

The threshold potential is defined as the first decreasing zero of the total conductance curve:
\begin{equation}\label{eq:v_th_eq_si}
    g_\text{t}(\Vth) = 0\text{~ with ~} g_\text{t}(\Vth - \delta V) > 0 > g_\text{t}(\Vth + \delta V), \quad\text{for all sufficiently small $\delta V > 0$}\quad.
\end{equation}

\begin{figure}[H]
    \centering
    \includegraphics[width=0.5\linewidth]{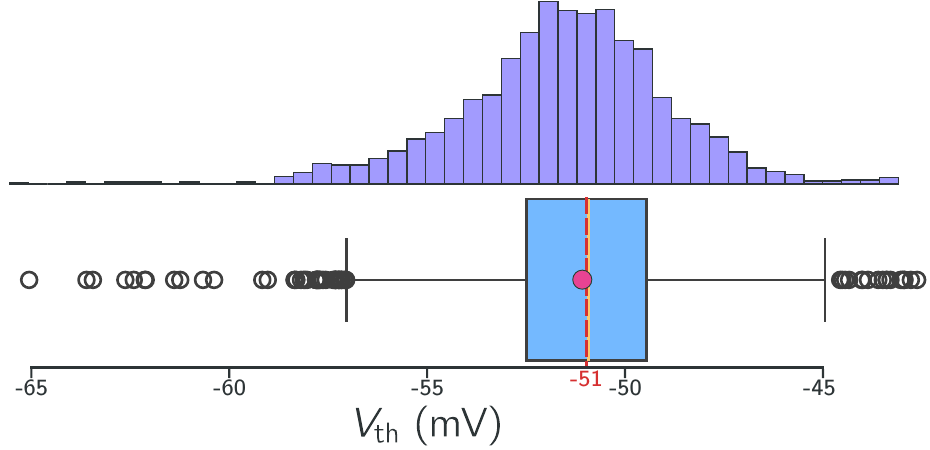}
    \caption{\textbf{Histogram of threshold voltage estimates ($\Vth$) for 8,000 STG model instances.} The distribution remains nearly identical to that obtained with 4,000 instances, indicating convergence around $\Vth \approx \SI{-51}{\milli\volt}$. The mean (pink dot) and median (yellow line) are close to the estimated value (red line).}
    \label{fig:stg_vth}
\end{figure}

To justify the shared-value approximation, we sampled 4,000 conductance vectors from $\mathcal{D}_\text{analysis}$ (see Section~\ref{sec:conductance_distributions}) and computed individual threshold voltages using Eq.~\ref{eq:v_th_eq_si} via a bisection method. The resulting distributions are narrow and unimodal, yielding $\Vth \approx \SI{-51}{\milli\volt}$ for the STG model and $\Vth \approx \SI{-55.5}{\milli\volt}$ for the DA model. These estimates are robust to sample size: repeating the analysis with 8,000 instances produced nearly identical values. Figure~\ref{fig:stg_vth} shows the histogram obtained for 8,000 STG instances, confirming that most threshold values cluster tightly around the estimate.

\subsection{Conductance distributions}
\label{sec:conductance_distributions}

In practice, we use two distinct distributions of maximal conductances, denoted $\bar{g} \sim \mathcal{D}$, each serving a different purpose:
\begin{itemize}
    \item $\mathcal{D}_\text{analysis}$: a broad distribution covering a wide region of the conductance space, used for preliminary analyses (threshold voltage estimation, DIC space exploration) prior to population generation.
    \item $\mathcal{D}_\text{generation}$: a more concentrated distribution, specifically employed during generation of degenerate populations.
\end{itemize}

The rationale for using two distributions stems from their different objectives. The distribution $\mathcal{D}_\text{analysis}$ is designed to explore a wide range of instances and observe various properties of the CBM under study, requiring broad coverage to avoid losing information. The distribution $\mathcal{D}_\text{generation}$ defines a parameter space that leads to degenerate behaviors: it should be spread enough to allow for significant degeneracy while minimizing the impact on firing variability. The variability in firing patterns when enforcing DIC values at threshold depends on the distribution used for uncompensated conductances. Generated instances should thus exhibit similar activity to be considered degenerate ($\mathcal{D}_\text{generation}$ concentrated), while maintaining large variability in conductance values ($\mathcal{D}_\text{generation}$ spread). In essence, $\mathcal{D}_\text{generation}$ should maximize spread to facilitate degeneracy, yet remain concentrated enough to ensure similar activity within each population when DIC values are enforced.

\subsubsection*{Leak conductance distribution}

In both distributions, we adopted a $\GGamma$ distribution for the leak conductance, rather than the uniform distribution used in~\cite{fyonDimensionalityReductionNeuronal2024}. The $\GGamma$ distribution models variables with positive support and does not impose a hard upper bound. With suitable parameters, it also stabilizes normalization by minimizing density near zero. The shape and scale parameters are chosen to match the first and second moments of the corresponding uniform distribution from~\cite{fyonDimensionalityReductionNeuronal2024}. The probability density function is:
\begin{equation}
    X \sim \GGamma(k, \theta) \implies p_{X}(x) = \dfrac{1}{\Gamma(k)\theta^k}x^{k-1}e^{-\frac{x}{\theta}}\quad,
\end{equation}
where $k$ is the shape parameter and $\theta$ is the scale parameter.

\subsubsection*{Homogeneous scaling}

In $\mathcal{D}_\text{generation}$, conductances are not independent: the leak conductance is sampled first, and the remaining conductances are scaled by the factor $\tfrac{g_\leak}{g_{\leak, \text{mean}}}$. This controls the effect of homogeneous scaling~\cite{fyonDimensionalityReductionNeuronal2024}, ensuring that relative conductance ratios are preserved across different leak values.

\subsubsection*{STG model distributions}

Table~\ref{table:stg_distributions} reports both distributions for the STG model. In $\mathcal{D}_\text{generation}$, no explicit bounds are set for $\bar{g}_\Na$, $\bar{g}_\Kd$, and $\bar{g}_\Hh$, since these are compensated during the first step of the generation procedure (Algorithm~\ref{algo:cbm_compensation}).

\begin{table}[!htbp]

    \caption[Distributions of maximal conductances in the STG model.]{\textbf{Distributions of maximal conductances in the STG model.} Two distinct distributions are used at different stages: (a) $\mathcal{D}_\text{analysis}$, a broad distribution for preliminary analyses, and (b) $\mathcal{D}_\text{generation}$, used at inference and generation. Values are in $\si{\mScmsq}$.}
    \label{table:stg_distributions}
    
    \centering
    
    \begin{subtable}{\textwidth}
    
        \caption{Preliminary distribution $\mathcal{D}_\text{analysis}$ for broad exploratory analyses. Conductances are sampled independently from uniform distributions with predefined upper bounds, except for the leak conductance, which follows a gamma distribution.}
        \label{table:naive_conductances_ranges}
        
        \centering
        \begin{tabular}{lccccccc}
            \toprule
            Conductance $\bar{g}_i \sim \mathcal{U}(0;\bar{g}_{\max})$ & $\bar{g}_\Na$ & $\bar{g}_\Kd$ & $\bar{g}_\CaT$ & $\bar{g}_\CaS$ & $\bar{g}_\KCa$ & $\bar{g}_\A$ & $\bar{g}_\Hh$ \\
            \midrule
            Maximum value $\bar{g}_{\max}$ & 8000 & 350 & 12 & 50 & 250 & 600 & 0.7 \\
            \bottomrule
        \end{tabular}
        
        \vspace{0.25cm}
        
        \begin{tabular}{lcc}
            \toprule
            Conductance $\bar{g}_i \sim \GGamma(k,\theta)$ & $k$ & $\theta$ \\
            \midrule
            $g_\leak$ & 3 & $\frac{1}{300}$ \\
            \bottomrule
        \end{tabular}
        
    \end{subtable}

    \vspace{0.5cm}

    \begin{subtable}{\textwidth}
    
        \caption{Generation distribution $\mathcal{D}_\text{generation}$ for producing degenerate populations. This distribution is more concentrated and incorporates homogeneous scaling by normalizing conductances relative to the leak conductance.}
        \label{table:conductances_ranges}
        
        \centering
        \begin{tabular}{lccccccc}
            \toprule
            Conductance $\bar{g}_i \sim \mathcal{U}(\bar{g}_{\min};\bar{g}_{\max})\frac{g_\leak}{g_{\leak, \text{mean}}}$ & $\bar{g}_\Na$ & $\bar{g}_\Kd$ & $\bar{g}_\CaT$ & $\bar{g}_\CaS$ & $\bar{g}_\KCa$ & $\bar{g}_\A$ & $\bar{g}_\Hh$ \\
            \midrule
            Minimum value $\bar{g}_{\min}$ & --- & 70 & 2 & 6 & 140 & --- & --- \\
            Maximum value $\bar{g}_{\max}$ & --- & 140 & 7 & 22 & 180 & --- & --- \\
            \bottomrule
        \end{tabular}
        
        \vspace{0.25cm}
        
        \begin{tabular}{lcc}
            \toprule
            Conductance $\bar{g}_i \sim \GGamma(k,\theta)$ & $k$ & $\theta$ \\
            \midrule
            $g_\leak$ & 27 & $\frac{1}{2570}$ \\
            \bottomrule
        \end{tabular}
        
    \end{subtable}

\end{table}

\subsubsection*{DA model distributions}

A similar approach was followed for the DA model, reported in Table~\ref{table:da_distributions}. The only distinction is that $\bar{g}_\NMDA$ was set to be directly proportional to the leak conductance, corresponding to a fixed biological network around the cell across instances, as this current is mainly involved in synaptic transmission.

\begin{table}[!htbp]
    \caption[Distributions of maximal conductances in the DA model.]{\textbf{Distributions of maximal conductances in the DA model.} Two distinct distributions are used: (a) $\mathcal{D}_\text{DA-analysis}$ for preliminary analyses, and (b) $\mathcal{D}_\text{DA-generation}$ for inference and generation. Values are in $\si{\mScmsq}$.}
    \label{table:da_distributions}
    
    \centering

    \begin{subtable}{\textwidth}
    
        \caption{Preliminary distribution $\mathcal{D}_\text{DA-analysis}$ for exploratory analyses.}        
        \label{table:DA_naive_conductances_ranges}
        \centering
        \begin{tabular}{lcccccc}
            \toprule
            Conductance $\bar{g}_i \sim \mathcal{U}(0;\bar{g}_{\max})$ & $\bar{g}_\Na$ & $\bar{g}_\Kd$ & $\bar{g}_\CaL$ & $\bar{g}_\CaN$ & $\bar{g}_\ERG$ & $\bar{g}_\text{NMDA}$ \\
            \midrule
            Maximum value $\bar{g}_{\max}$ & 60 & 20 & 0.1 & 0.12 & 0.25 & 0.012 \\
            \bottomrule
        \end{tabular}
        
        \vspace{0.25cm}
        
        \begin{tabular}{lcc}
            \toprule
            Conductance $\bar{g}_i \sim \GGamma(k,\theta)$ & $k$ & $\theta$ \\
            \midrule
            $\bar{g}_\text{leak}$ & 3 & $\frac{1}{300}$ \\
            \bottomrule
        \end{tabular}
        
    \end{subtable}

    \vspace{0.5cm}

    \begin{subtable}{\textwidth}
    
        \caption{Generation distribution $\mathcal{D}_\text{DA-generation}$ for producing degenerate populations.}        
        \label{table:DA_conductances_ranges}
        \centering
        \begin{tabular}{lccccccc}
            \toprule
            Conductance $\bar{g}_i \sim \mathcal{U}(\bar{g}_{\min};\bar{g}_{\max})\frac{g_\text{leak}}{\bar{g}_{\text{leak, mean}}}$ & $\bar{g}_\Na$ & $\bar{g}_\Kd$ & $\bar{g}_\CaL$ & $\bar{g}_\CaN$ & $\bar{g}_\ERG$ & $\bar{g}_\text{NMDA}$ \\
            \midrule
            Minimum value $\bar{g}_{\min}$ & --- & 6 & 0.015 & --- & --- & 0.012 \\
            Maximum value $\bar{g}_{\max}$ & --- & 10 & 0.075 & --- & --- & 0.012 \\
            \bottomrule
        \end{tabular}
        
        \vspace{0.25cm}
        
        \begin{tabular}{lcc}
            \toprule
            Conductance $\bar{g}_i \sim \GGamma(k,\theta)$ & $k$ & $\theta$ \\
            \midrule
            $g_\leak$ & 28.76 & $\frac{1}{2238}$ \\
            \bottomrule
        \end{tabular}
        
    \end{subtable}

\end{table}

\subsection{Algorithmic procedure}

We summarize the full procedure for generating populations of CBMs targeting specific DIC values as an algorithmic workflow (Algorithm~\ref{algo:cbm_compensation}). The method applies to a generic $N$-channel neuron model and consists of two main steps: generating a spontaneously active population and modulating it to reach a target point in the DIC space.

\begin{algorithm}[!htbp]
\caption{Iterative compensation procedure for CBM population generation.}
\label{algo:cbm_compensation}
\begin{algorithmic}[1]

\State \textbf{Input:} Number of channels $N$, DIC target $g_\text{DICs} = (g_\s, g_\us)$, threshold value $\Vth$, conductance distributions $\mathcal{D}_\text{generation}$ and the leak distribution $\GGamma$.
\State \textbf{Output:} Population of CBM instances with compensated conductances.

\Procedure{GenerateSpontaneousPopulation}{}
    \State Draw leak conductance $g_\leak \sim \GGamma$.
    \State Draw $N-3$ maximal conductances from $\mathcal{D}_\text{generation}$ scaled proportionally to $g_\leak$.
    \State Compute the remaining 3 compensated conductances to impose a sufficiently negative $g_\f(\Vth)$:
    \begin{itemize}
        \item STG: $\bar{g}_\text{comp., spont.} = (\bar{g}_\Na, \bar{g}_\Kd, \bar{g}_\Hh)$
        \item DA: $\bar{g}_\text{comp., spont.} = (\bar{g}_\Na, \bar{g}_\CaN, \bar{g}_\ERG)$
    \end{itemize}
    \State Target the DIC values in Table~\ref{table:dic_targets_step1}.
\EndProcedure

\Procedure{ModulatePopulationToTargetDIC}{population, $g_\text{DICs}$}
    \State Initialize iteration counter $k = 0$
    \While{$k < K_\text{max}$} \Comment{e.g., $K_\text{max}=5$}
        \State Update compensated conductances using the iterative solver based on $S(\bar{g}_\text{comp.})$ \\ \hspace{2.7em} to approach $g_\text{DICs}$.
        \State Increment $k \gets k+1$
    \EndWhile
    \State Remove any instances with negative compensated conductances.
    \State Select conductances to compensate as:
    \begin{equation}
        \bar{g}_\text{comp.} = 
        \begin{cases}
            (\bar{g}_\CaS, \bar{g}_\Hh), & g_\s < 0\\
            (\bar{g}_\A, \bar{g}_\Hh), & g_\s > 0
        \end{cases}\quad, \quad
        \bar{g}_\text{DA-comp.} = 
        \begin{cases}
            (\bar{g}_\text{ERG}, \bar{g}_\text{CaL}), & g_\s < 0\\
            (\bar{g}_\text{ERG}, \bar{g}_\text{Kd}), & g_\s > 0
        \end{cases}
    \end{equation}
\EndProcedure

\State \textbf{Execute:}
\State $population \gets \text{GenerateSpontaneousPopulation}()$
\State $population \gets \text{ModulatePopulationToTargetDIC}(population, g_\text{DICs})$
\State \textbf{return} $population$

\end{algorithmic}
\end{algorithm}

The first step produces a spontaneously active (spiking) population, while the second step iteratively adjusts the conductances to reach the target DIC values $(g_\s, g_\us)$. We note that some negative conductances may remain in the DA model for $g_\us < 1.5$, but the regime-dependent selection of conductances for compensation reduces most invalid instances.

\subsection{Synthetic dataset generation}
\label{sec:dataset_generation}

We constructed a large open-source synthetic dataset spanning a broad range of DIC values and corresponding spike trains~\cite{brandoit2025degenerate}. Instead of sampling conductance parameters directly, we uniformly sampled the slow and ultra-slow DICs $g^* = \left(g_\s(\Vth);~g_\us(\Vth)\right)$, since these components primarily shape firing activity. The fast DIC $g_\f(\Vth)$ was constrained qualitatively to ensure spontaneous activity (sufficiently negative).

\subsubsection*{Sampling bounds}

The bounds on the $(g_\s, g_\us)$ space were derived by extensively sampling the $\mathcal{D}_\text{analysis}$ distributions. Based on 4,000 instances (then 8,000 to ensure convergence) used to estimate the threshold voltage, we extracted the observed DIC bounds at threshold. For the STG model, the DIC values were largely enclosed in $(g_\s(\Vth);~ g_\us(\Vth)) \in [-20,~20]\times[-2,~20]$. We restricted the ultra-slow DIC to be positive (effectively using $[-20,~20]\times[0,~20]$) because we observed generation issues for negative values: populations were not degenerate, and such negative ultra-slow threshold DIC values corresponded to the negative outliers in Fig~\ref{fig:stg_vth}. We hypothesize that the shared threshold voltage approximation was not appropriate for such instances. For the DA model, the bounds were $[-15,~15]\times[0,~20]$, restricted during inference to $g_\s \in [-10,~15]$ as all instances generated from $g_\s < -10$ were silent.

\subsubsection*{Dataset construction}

For each of the $N = 75{,}000$ sampled DIC pairs, we generated a degenerate population of $M = 16$ CBM instances using the iterative compensation algorithm, yielding a total of $|\mathcal{T}| = 1{,}200{,}000$ simulated neurons. Each instance was simulated under noisy current injection as described in Section~\ref{sec:cbm_models}. The total simulation duration was $\SI{5000}{\milli\second}$ for the STG model and $\SI{12000}{\milli\second}$ for the DA model. From each simulated trace, only spike times were retained:
\begin{equation*}
V(t) \xrightarrow[]{\text{transformed into}} ~x = [t_1, t_2, \dots, t_{N_\text{spikes}}], \quad t_1 < t_2 < \dots < t_{N_\text{spikes}}\quad.
\end{equation*}

The final STG dataset consisted of 51.48\% spiking and 48.28\% bursting neurons, with 0.24\% silent instances discarded. To prevent data leakage, instances from the same population were kept together and never split across partitions: $|\mathcal{T}_\text{train}| = 1{,}000{,}000$ instances for training, $|\mathcal{T}_\text{val}| = 200{,}000$ for validation, and $200{,}000$ for testing. The test set was generated only after the architecture was fully trained and fixed, ensuring that performance metrics reflect generalization to unseen DIC targets and conductance configurations.

For the DA model, a reduced dataset of approximately 40\% of the STG size was used, since transfer learning via LoRA adapters facilitates model adaptation with smaller datasets. The number of sampled configurations was set to yield approximately 25,000 active (non-silent) populations in total.

\subsection{Target DIC values}

Table~\ref{tab:dic_targets_summary} reports the exact DIC values targeted for the various results presented in this work.

The residual analysis of Fig~\ref{fig:fig_ICA_SI} was based on sampling within the STG range reported above. We generated 5,000 populations of 250 instances from uniform sampling of the DIC targets and compensated $\bar{g}_\CaS$ and $\bar{g}_\A$. After discarding populations with negative conductance values, 1,633 remained for the analysis. The same random set of values was shared across all methods, so the analysis directly compared the results of the compensation step.

\begin{table}[!htbp]
    \caption{\textbf{Summary of target DIC values used in this work.}}
    \label{tab:dic_targets_summary}
    \centering
    \begin{subtable}{\textwidth}
        \caption{\textbf{Target DIC values at threshold for the first step of the generation procedure.} These correspond to spontaneously active spiking populations.}
        \label{table:dic_targets_step1}
        \centering
        \begin{tabular}{lccc}
            \toprule
            Target DIC values & $g_\f$ &$g_\s$ & $g_\us$ \\
            \midrule
            STG & -6.2 & 4 & 5 \\
            DA & -12.95 & 0.5 & 5 \\
            \bottomrule
        \end{tabular}
    \end{subtable}

    \vspace{0.3cm}

    \begin{subtable}{\textwidth}
        \caption{\textbf{Target DIC values for the example populations of Fig.~\ref{fig:fig_ICA_SI}}}
        \label{table:dic_fig2_targets}
        \centering
        \begin{tabular}{lcc}
            \toprule
            Target DIC values & $g_\s$ & $g_\us$ \\
            \midrule
            Spiking & 5 & 4 \\
            Bursting & -2.71 & 5.63 \\
            \bottomrule
        \end{tabular}
    \end{subtable}
\end{table}

\subsection{Spike definition and firing activity descriptors}

Throughout the paper, we rely on descriptors of neuronal activity. These descriptors are \textbf{not} inputs to the deep learning architecture; however, they provide useful intermediates to characterize the firing pattern of individual instances (see Fig.~\ref{fig:fig_ICA_SI} and Fig.~6 from the main text).

\subsubsection*{Spike extraction}

When simulating CBMs, we obtained the full voltage trace $V(t)$. To simulate experimental conditions where only spike timing is recorded, we extracted spike times from $V(t)$ using a two-threshold method. A spike was considered to occur at the midpoint between the voltage crossing an upper threshold $V_\mathrm{up} = \SI{10}{\milli\volt}$ from below and subsequently crossing a lower threshold $V_\mathrm{down} = \SI{0}{\milli\volt}$ from above. Formally, for each spike $i$:
\[
t_i = \frac{t_\mathrm{up}^{(i)} + t_\mathrm{down}^{(i)}}{2}\quad,
\]
where $t_\mathrm{up}^{(i)}$ is the time when $V(t)$ first exceeds $V_\mathrm{up}$ and $t_\mathrm{down}^{(i)}$ is the time when it next falls below $V_\mathrm{down}$. Applying this procedure to all spikes produces the sequence:
\[
V(t) \;\xrightarrow{\text{spike extraction}}\; x = [t_1, t_2, \dots, t_{N_\mathrm{spikes}}]\quad.
\]

\subsubsection*{Activity descriptors}

Given a sequence of spike times $x = [t_1, \dots, t_{N_\text{spikes}}]$ and the corresponding inter-spike intervals $x_\text{ISI} = \Delta x = [t_2-t_1, t_3-t_2, \dots, t_{N_\text{spikes}} - t_{N_\text{spikes} -1}]$, we computed the following descriptors.

\paragraph{Spiking activity.} We reported the mean firing frequency:
\begin{equation}
    f_\text{spk} = \left(\dfrac{1}{N_\text{spikes} - 1} \sum_{i = 1}^{N_\text{spikes} - 1}\Delta x_i\right)^{-1}\quad.
\end{equation}

\paragraph{Bursting activity.} We computed metrics based on burst extraction. From the spike time sequence, we constructed a sequence of bursts $\mathcal{B} = [B_1, \dots, B_{N_\text{burst}}]$, where each $B_i$ is a sequence of $N_{i, \text{spikes}}$ consecutive spikes, split based on the mid-range value of $x_\text{ISI}$. The first spike of each burst was such that the duration since the previous spike exceeded $\tfrac{\min(x_\text{ISI}) + \max(x_\text{ISI})}{2}$. The first and last bursts were discarded, as they could be incomplete due to simulation boundaries. We then reported:

\begin{enumerate}
    \item The mean intra-burst frequency:
    \begin{equation}
        f_\text{intra} = \left(\dfrac{1}{N_\text{burst}} \sum_{i = 1}^{N_{i,\text{spikes}} - 1}\Delta B_i\right)^{-1}\quad.
    \end{equation}
    \item The mean inter-burst frequency, i.e., the inverse of the mean duration between bursts:
    \begin{equation}
        f_\text{inter} = \left(\dfrac{1}{N_\text{burst} - 1} \sum_{i = 1}^{N_\text{burst} - 1}(B_{i+1})_{N_{i+1,\text{spikes}}} - (B_{i})_{1}\right)^{-1}\quad.
    \end{equation}
    \item The mean burst duration:
    \begin{equation}
        \text{Duration} = \dfrac{1}{N_\text{burst}} \sum_{i=1}^{N_\text{burst}} \sum_{j=1}^{N_{i,\text{spikes}} - 1}(\Delta B_{i})_j\quad.
    \end{equation}
    \item The mean number of spikes per burst:
    \begin{equation}
        \# = \dfrac{1}{N_\text{burst}} \sum_{i=1}^{N_\text{burst}} N_{i, \text{spikes}}\quad.
    \end{equation}
    Some bursting neurons were slightly irregular (for example, producing alternating bursts of 3 and 4 spikes), and we kept the mean as a non-integer value in those cases. This is why this metric can take non-integer values despite the quantized nature of spikes per burst.
\end{enumerate}

\section{Heterogeneous populations at the spiking-bursting transition}
\label{sec:heterogeneous_transition}

Throughout the paper, figures such as Fig.~2 from the main text show an overlap between the DIC region associated with bursting and the one associated with spiking. The transition between regimes is roughly located at $g_\s \approx 0$, with $g_\s > 0$ corresponding to spiking and $g_\s < 0$ to bursting, and almost no dependency on $g_\us$. However, in practice, this transition is not sharp, and a whole region marks it. In this region, generated populations are heterogeneous, containing a mix of spiking and bursting instances. The bursting instances in this transition zone typically produce doublets (bursts with only 2 spikes) or alternations of single spikes and doublets.

\begin{figure}[H]
    \centering
    \includegraphics[width=0.85\linewidth]{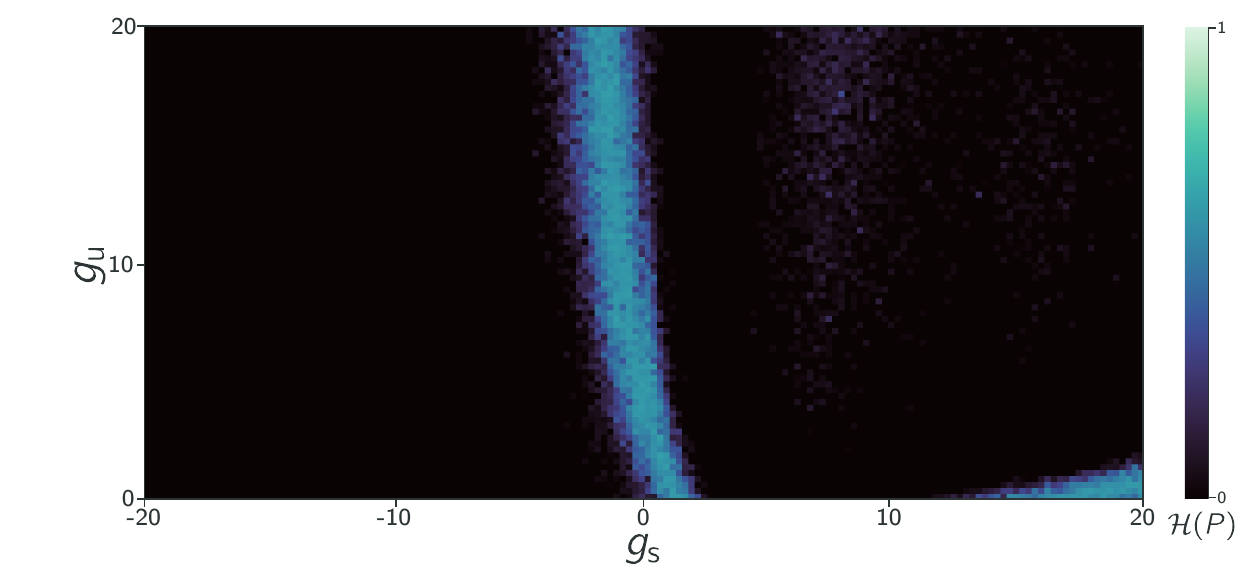}
    \caption{\textbf{Population heterogeneity across the STG DIC space.} The spiking-bursting transition ($g_\s \approx 0$) corresponds to a highly heterogeneous region where populations contain a mix of spiking and bursting instances.}
    \label{fig:entropy}
\end{figure}

To characterize this region, we generated populations across the DIC space and computed the class entropy of each population $P$:
\begin{equation}
\mathcal{H}(P) = - \sum_{t \in \{\text{silent}, \text{spiking}, \text{bursting}\}} p_t \log_3(p_t) \in [0, 1]\quad,
\end{equation}
where homogeneous populations have zero entropy and highly heterogeneous populations have an entropy of one. Figure~\ref{fig:entropy} highlights the spiking-bursting transition based on this entropy measure. A small region of mixed spiking-silent instances is also visible in the bottom right of the DIC space.

This analysis directly addresses the question of what proportion of generated models preserve the nominal activity type. Outside the narrow transition zone near $g_\s \approx 0$, populations are homogeneous (entropy $\approx 0$), meaning that virtually all instances preserve the expected activity type: spiking for $g_\s > 0$ and bursting for $g_\s < 0$. The mixed populations observed at the transition are not a failure of the method but rather reflect a genuine dynamical property: near the bifurcation between regimes, the system is sensitive to small parameter variations, and individual instances within a degenerate population may fall on either side of the boundary. Silent neurons are rare throughout the DIC space, representing less than 0.24\% of all generated instances, and occur primarily at the extreme boundaries where DIC constraints approach non-physiological values.

\section{Deep learning architecture and training}

\subsection{Architecture overview}

\begin{figure}[H]
  \centering
  \includegraphics[width=\textwidth]{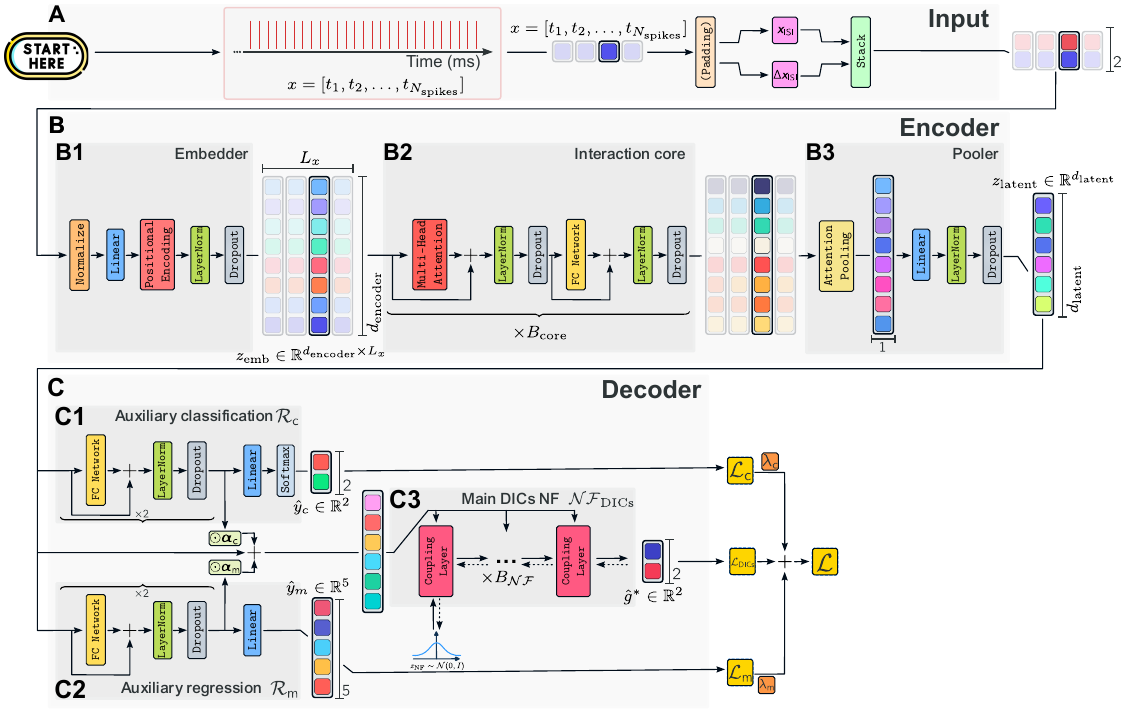}
  \caption{{\bf The deep learning architecture.}
\textbf{(A)} The input to the model consists of spike time sequences, from which ISIs and delta ISIs are extracted. These features are then stacked and fed into the encoder. \textbf{(B)} The encoder processes the input through three main components: the embedder, the interaction core, and the pooler. The embedder transforms the input sequence into a normalized, higher-dimensional representation. The interaction core processes this representation using multi-head attention mechanisms and fully-connected networks. The pooler aggregates the variable-length representation into a fixed-size latent representation. \textbf{(C)} The decoder transforms the fixed-size latent representation into a conditional density over DICs via a normalizing flow, and includes auxiliary heads for classifying neuronal activity and regressing electrical activity metrics.}
  \label{fig:fig_architecture}
\end{figure}

The architecture (Fig~\ref{fig:fig_architecture}) comprises an attention-based encoder~\cite{vaswaniAttentionAllYou2017} and a multi-headed decoder. The encoder maps variable-length spike time sequences to a fixed-size latent vector $z_\text{latent} \in \mathbb{R}^{d_\text{latent}}$, and the decoder processes this representation through three parallel heads to produce both the primary output and auxiliary predictions that regularize training.

\paragraph{Encoder.} Given a spike train $x$, we compute inter-spike intervals (ISIs) and their second-order differences to enhance burst detection. A logarithmic transform $\log(1 + x_\text{ISI})$ is applied to stabilize the ISI distribution, and the resulting features are standardized using training set statistics. The embedder (Fig~\ref{fig:fig_architecture}B1) projects these normalized features into a higher-dimensional space enriched with sinusoidal positional encoding~\cite{vaswaniAttentionAllYou2017}. The interaction core (Fig~\ref{fig:fig_architecture}B2) consists of $B_\text{core}$ stacked transformer blocks, each combining multi-head self-attention with a position-wise fully connected network using GELU activations~\cite{hendrycksGaussianErrorLinear2023}, residual connections, layer normalization~\cite{baLayerNormalization2016}, and dropout. A self-attention pooling layer (Fig~\ref{fig:fig_architecture}B3) aggregates the variable-length output into a fixed-size latent vector $z_\text{latent} \in \mathbb{R}^{d_\text{latent}}$.

\paragraph{Decoder.} The decoder comprises three heads operating on the shared latent representation. The primary head is a RealNVP-style normalizing flow~\cite{dinh2017densityestimationusingreal} ($\mathcal{NF}_\text{DICs}$, Fig~\ref{fig:fig_architecture}C3) that models the conditional density $p_\theta(g^* \mid z_\text{latent})$ over DIC targets using stacked affine coupling layers, each conditioned on the latent vector. Two auxiliary heads are used only during training: a classification head ($\mathcal{R}_\text{c}$, Fig~\ref{fig:fig_architecture}C1) that predicts the firing regime (spiking or bursting), and a regression head ($\mathcal{R}_\text{m}$, Fig~\ref{fig:fig_architecture}C2) that predicts five activity descriptors (mean firing rate for spiking; intra-burst frequency, inter-burst frequency, burst duration, and spikes per burst for bursting). The auxiliary latent outputs are integrated with the encoder output via learnable element-wise mixing before conditioning the normalizing flow. At inference time, only the normalizing flow head is used.

\subsection{Training procedure}

The model is trained end-to-end by minimizing a composite loss $\mathcal{L} = \mathcal{L}_\text{flow} + \lambda_\text{m} \mathcal{L}_\text{m} + \lambda_\text{c} \mathcal{L}_\text{c}$, combining the negative log-likelihood of the normalizing flow ($\mathcal{L}_\text{flow}$) with a masked mean squared error on the activity descriptors ($\mathcal{L}_\text{m}$, where the mask selects regime-appropriate metrics) and a balanced cross-entropy for firing regime classification ($\mathcal{L}_\text{c}$). The weights $\lambda_\text{m}$ and $\lambda_\text{c}$ control the relative importance of the auxiliary tasks.

Three forms of data augmentation are applied during training: random cropping of spike trains to a window $D \sim \mathcal{U}[N_\text{spikes}/2, N_\text{spikes}]$, Gaussian jitter $\epsilon_i \sim \mathcal{N}(0, (2\,\text{ms})^2)$ on spike times, and 5\% spike dropout. These are applied independently per sample at each update.

Hyperparameters were optimized via random search over 100 configurations~\cite{bergstraRandomSearchHyperparameter2012}, using a lightweight pointwise regression decoder in place of the normalizing flow to reduce computational cost. Table~\ref{table:hyper_table} lists the explored hyperparameters and their distributions, and Table~\ref{table:final_hyperparams} reports the selected values. The selected configuration was then used to train the full model with the normalizing flow head. We used AdamW~\cite{loshchilovDecoupledWeightDecay2019} with a cosine annealing schedule with warm restarts (period $T_i = 10$ epochs, minimum learning rate $\eta_\text{min} = \eta / 10$). Training ran for 200 epochs, with validation every quarter epoch; the checkpoint with the lowest validation loss $\mathcal{L}_\text{flow}$ was retained. The final architecture comprises $150{,}572$ trainable parameters.

\begin{table}[!htbp]
\caption[Hyperparameters and distributions explored in random search.]{\textbf{Hyperparameters and distributions explored in random search.}
$\tilde{\lambda}_\text{c}$ and $\tilde{\lambda}_\text{m}$ are relative weights of the auxiliary losses to the primary DICs loss.}
    \label{table:hyper_table}
    \centering
    \renewcommand{\arraystretch}{1.1}
    \begin{tabular}{l|c|c}
    \toprule
        \textbf{Hyperparameter} & \textbf{Type / Distribution} & \textbf{Values or Range} \\
    \midrule
        Learning rate ($\eta$) & Log-uniform & $[10^{-5}, 5 \cdot 10^{-3}]$ \\
        Dropout ($p_\text{dropout}$) & Uniform & $[0.0, 0.4]$ \\
        Latent space dimension ($d_\text{latent}$) & Discrete & $\{16, 32, 64, 128\}$ \\
        Encoder space dimension ($d_\text{encoder}$) & Discrete & $\{16, 32, 64, 128\}$ \\
        Number of heads ($H$) & Discrete & $\{2, 4, 8\}$ \\
        Number of encoder blocks ($B_\text{core}$) & Discrete & $\{1, 2, 3, 4, 5, 6, 7, 8\}$ \\
        Number of decoder blocks ($B_{\mathcal{R}_\text{DICs}}$) & Discrete & $\{2, 4, 6, 8, 10\}$ \\
        Activation function ($\sigma(\cdot)$) & Categorical & \texttt{relu}, \texttt{gelu}, \texttt{silu}, \texttt{tanh} \\
        Apply log transform to input? & Boolean & \texttt{true}, \texttt{false} \\
        Regression weighting factor ($\tilde{\lambda}_\text{m}$) & Uniform & $[0.01, 10.0]$ \\
        Classification weighting factor ($\tilde{\lambda}_\text{c}$) & Uniform & $[0.01, 10.0]$ \\
        Batch size ($|B|$) & Discrete & $\{16, 32, 64, 128\}$ \\
    \bottomrule
    \end{tabular}
\end{table}

\begin{table}[!htbp]
    \caption{\textbf{Final hyperparameters for the STG backbone.} Values selected after random search, yielding the best validation performance.}
    \label{table:final_hyperparams}
    \centering
    \renewcommand{\arraystretch}{1.1}
    \begin{tabular}{l|c}
        \toprule
        \textbf{Hyperparameter} & \textbf{Final value} \\
        \midrule
        Learning rate ($\eta$) & $2.10 \times 10^{-5}$ \\
        Dropout ($p_\text{dropout}$) & $0.034$ \\
        Latent space dimension ($d_\text{latent}$) & $16$ \\
        Encoder space dimension ($d_\text{encoder}$) & $64$ \\
        Number of heads ($H$) & $8$ \\
        Number of encoder blocks ($B_\text{core}$) & $4$ \\
        Number of decoder blocks ($B_{\mathcal{R}_\text{DICs}}$) & $2$ \\
        Activation function ($\sigma(\cdot)$) & \texttt{gelu} \\
        Apply log transform to input? & \texttt{true} \\
        Regression weighting factor ($\tilde{\lambda}_\text{m}$) & $0.0919$ \\
        Classification weighting factor ($\tilde{\lambda}_\text{c}$) & $5.44$ \\
        Batch size ($|B|$) & $32$ \\
        \bottomrule
    \end{tabular}
\end{table}

\subsection{Evaluation metrics}

During hyperparameter optimization, the primary selection metric was the MAE of the pointwise regression decoder on the validation set. At test time, we report auxiliary task performance: MAE on activity descriptors for the regression head, and balanced accuracy for the classification head. These metrics verify that the encoder captures meaningful temporal structure from raw spike trains. The quality of predicted DIC targets is ultimately assessed through the end-to-end posterior predictive checks described in the main text (Fig.~4 and Fig.~5).

\subsection{Posterior calibration diagnostics}
\label{sec:posterior_diagnostics}

Because our architecture learns a conditional density $p_\theta(g^* \mid x)$ rather than a point estimate, it is essential to verify that the learned posteriors are well calibrated. Following the recommendations of~\cite{hermans2022trustcrisissimulationbasedinference}, we assessed calibration using three complementary diagnostics: Tests of Accuracy with Random Points (TARP)~\cite{lemos2023samplingbasedaccuracytestingposterior}, simulation-based calibration (SBC) rank histograms~\cite{taltsValidatingBayesianInference2018}, and expected coverage tests~\cite{cook2006validation}. All diagnostics were evaluated on held-out validation data for both the STG and DA models. Results are summarized in Fig~\ref{fig:calibration_diagnostics}.

\subsubsection*{Tests of Accuracy with Random Points (TARP)}

TARP provides a joint calibration diagnostic sensitive to both coverage and bias. For each of $N$ validation inputs, $L = 2{,}000$ posterior samples are drawn from the learned conditional density. A set of $R = 1{,}000$ random reference points is sampled uniformly over the DIC parameter space. For each reference point, the test computes the fraction of posterior samples that lie closer (in Euclidean distance) to the reference than the true parameter value does. If the posterior is well calibrated, these fractions should be uniformly distributed and the resulting empirical coverage curve should closely follow the identity line. Deviations above the diagonal indicate conservative (overdispersed) posteriors, while deviations below indicate overconfident ones. We quantified departure from ideal calibration using the expected calibration error (ECE), defined as the mean absolute deviation between the empirical coverage curve and the diagonal. 95\,\% bootstrap confidence intervals were computed from 1{,}000 bootstrap resamples of the reference points.

For both models, the empirical TARP coverage curves closely follow the diagonal and remain within the 95\,\% confidence band across all credibility levels (Fig~\ref{fig:calibration_diagnostics}, first column). The ECE values are small for both models, indicating no detectable joint miscalibration.

\subsubsection*{Simulation-based calibration rank histograms}

SBC provides a marginal calibration check for each parameter dimension independently~\cite{taltsValidatingBayesianInference2018}. For each validation input, the rank of the true DIC value among the $L$ posterior samples is computed: $r = \sum_{l=1}^{L} \mathbb{I}[g_l < g^*]$. If the posterior is calibrated, the distribution of ranks should be uniform. A U-shaped histogram signals an overdispersed posterior, an inverted-U shape signals overconcentration, and a skewed histogram signals systematic bias. We assessed uniformity using a chi-squared goodness-of-fit test with 20 histogram bins.

For both models and both DIC dimensions, the rank distributions are approximately uniform and the chi-squared tests do not reject uniformity at the 5\,\% significance level (Fig~\ref{fig:calibration_diagnostics}, second and third columns), confirming that the marginal posteriors are well calibrated.

\subsubsection*{Expected coverage tests}

As a third diagnostic, we performed marginal expected coverage tests~\cite{cook2006validation}. For each credibility level $\alpha \in [0, 1]$, the symmetric $\alpha$-credible interval is constructed from the posterior samples for each validation input, and the empirical coverage is computed as the fraction of inputs for which the true value falls within this interval. We summarized calibration quality using the ECE and computed 95\,\% bootstrap confidence bands from 300 resamples.

\begin{figure}[H]
  \centering
  \includegraphics[width=\textwidth]{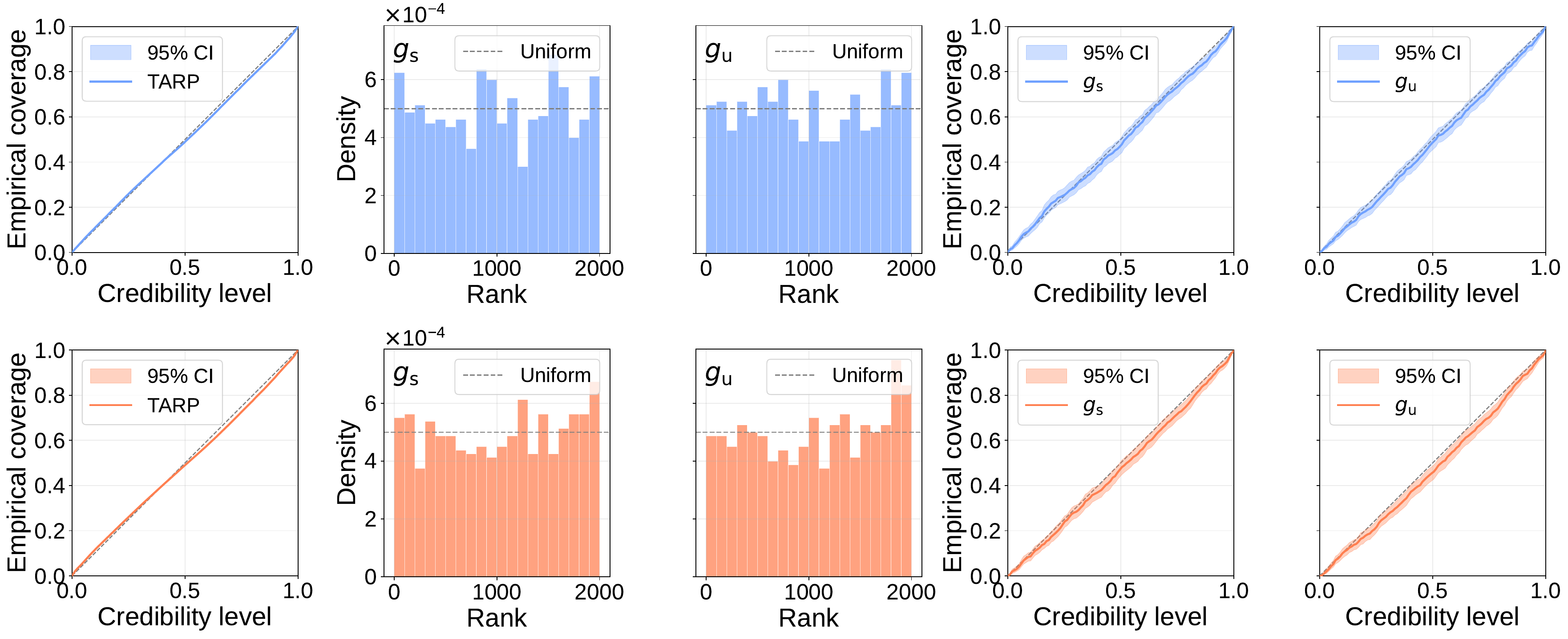}
  \caption{\textbf{Posterior calibration diagnostics for the STG (top, blue) and DA (bottom, orange) models.}
  From left to right: TARP coverage curves with 95\,\% bootstrap confidence bands; SBC rank histograms for $g_\text{s}(\Vth)$ and $g_\text{u}(\Vth)$ with the expected uniform density (dashed); and marginal expected coverage curves for $g_\text{s}(\Vth)$ and $g_\text{u}(\Vth)$ with 95\,\% bootstrap confidence bands.
  \textit{STG:} TARP ECE\,$=$\,$0.0104$; SBC $\chi^2$ $p$\,$=$\,$0.060$ ($g_\text{s}$), $0.493$ ($g_\text{u}$); expected coverage ECE\,$=$\,$0.0150$ ($g_\text{s}$), $0.0181$ ($g_\text{u}$); 90\,\% credible interval coverage\,$=$\,$0.874$ ($g_\text{s}$), $0.884$ ($g_\text{u}$).
  \textit{DA:} TARP ECE\,$=$\,$0.0140$; SBC $\chi^2$ $p$\,$=$\,$0.522$ ($g_\text{s}$), $0.178$ ($g_\text{u}$); expected coverage ECE\,$=$\,$0.0227$ ($g_\text{s}$), $0.0286$ ($g_\text{u}$); 90\,\% credible interval coverage\,$=$\,$0.875$ ($g_\text{s}$), $0.885$ ($g_\text{u}$).
  SBC $\chi^2$ tests do not reject the null hypothesis of uniform ranks at the 5\,\% significance level for either model or parameter; TARP and marginal expected coverage ECE values are small, indicating no meaningful deviation from calibration.}
  \label{fig:calibration_diagnostics}
\end{figure}

For both models and both dimensions, the empirical coverage closely tracks the diagonal and falls within the confidence band (Fig~\ref{fig:calibration_diagnostics}, fourth and fifth columns). The empirical coverage of the 90\,\% credible interval is close to the nominal level in all cases, confirming the absence of systematic over- or under-coverage.

Taken together, the three diagnostics reveal no evidence of posterior miscalibration for either model. These results provide strong evidence that the learned conditional density captures the structure of the posterior over DIC values, including its degeneracy, in a statistically reliable manner.

\subsection{Low-Rank Adaptation and transfer to the DA model}

\begin{figure}[!htbp]
  \centering
  \includegraphics[width=\textwidth]{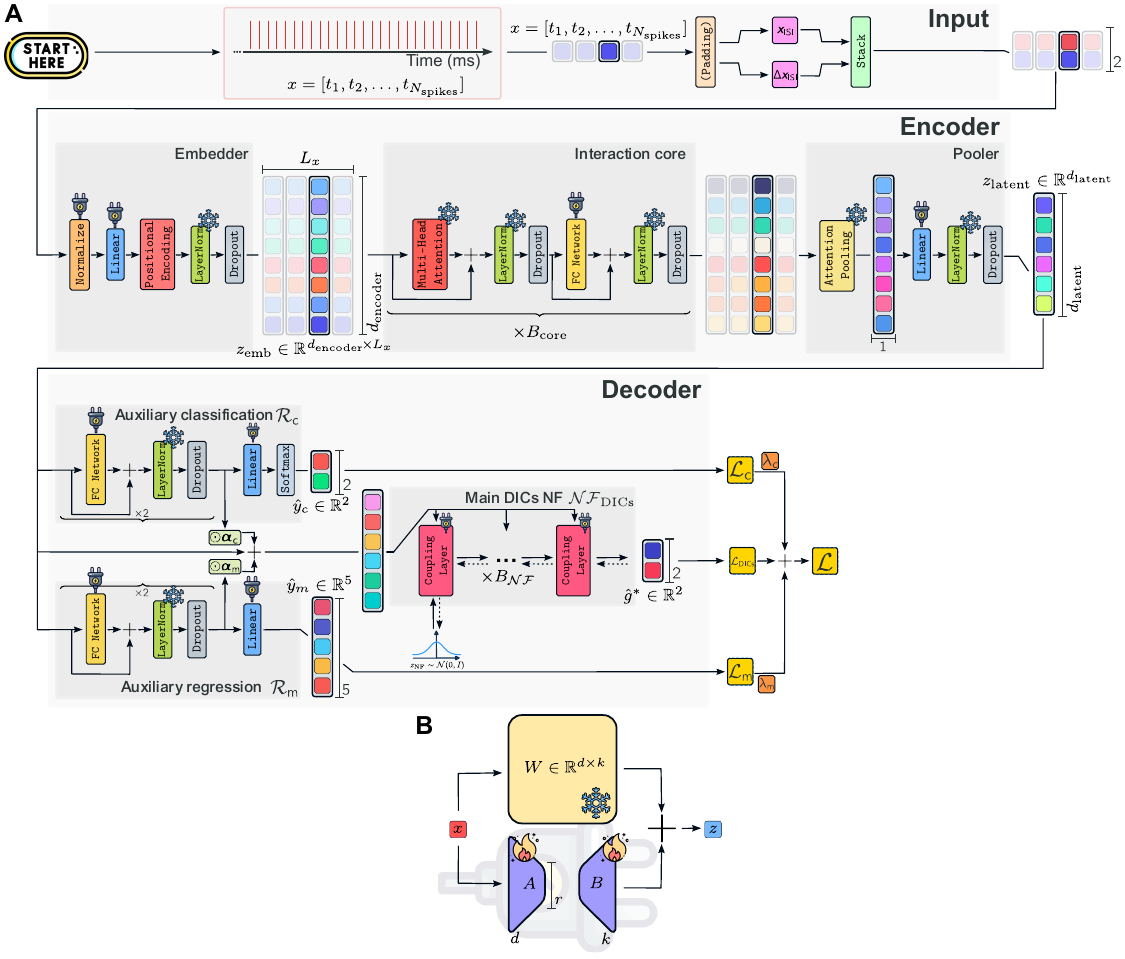}
  \caption{{\bf The deep learning architecture with LoRA to transfer from the STG to the DA.}
\textbf{(A)} The plug icon indicates the insertion of a LoRA adapter, where the original layer parameters are frozen and a new linear projection is added in an additive manner with new learnable parameters (panel (B)). The freeze icon signifies that the parameters of the corresponding layer are not retrained, preserving its original functionality and weights. The only exception is the normalization block (orange) at pipeline input, which is directly replaced by the new normalization statistics calculated on the DA train set. \textbf{(B)} The LoRA adapter introduces a small number of additional parameters through low-rank matrices $A \in \mathbb{R}^{d\times r}$ and $B \in \mathbb{R}^{r\times k}$. These matrices modify the matrix multiplication in fully connected layers by adding a learnable component to the frozen pre-trained weights $W \in \mathbb{R}^{d\times k}$.}
  \label{fig:fig_architecture_da}
\end{figure}

To adapt the pipeline to the DA neuron model, we applied parameter-efficient fine-tuning using Low-Rank Adaptation (LoRA)~\cite{huLoRALowRankAdaptation2021}. LoRA adapters were introduced in the fully connected layers of the network, while attention layers remained unmodified (Fig~\ref{fig:fig_architecture_da}A). The majority of backbone parameters were frozen, and only the newly introduced LoRA parameters were updated during training. The input normalization layer was recalculated based on the DA training set.

LoRA adapters rely on low-rank parameterized matrices ${A} \in \mathbb{R}^{d \times r}$ and ${B} \in \mathbb{R}^{r \times k}$ (Fig~\ref{fig:fig_architecture_da}B), which modify each weight matrix ${W} \in \mathbb{R}^{d \times k}$ as:
\begin{equation}
   {W}{x} \xrightarrow[\text{(plug icon)}]{\text{LoRA adapter}} \underbrace{{W}{x}}_{\text{frozen (freeze icon)}} + \underbrace{\dfrac{1}{r} {x}{A}{B}}_{\text{learnable (fire icon)}}.
\end{equation}

We set the rank to $r = 32$, which yields satisfactory transfer performance. This introduces additional parameters representing approximately 40\% of the total parameters that would be required for training a model from scratch. While the backbone is already lightweight ($150{,}572$ parameters), this result serves as a proof of concept that LoRA provides a practical strategy for extending the pipeline to new conductance-based models. In settings where the framework must support many distinct CBMs, each new model would only require storing a small set of LoRA weights on top of a shared backbone, making the approach well suited for scaling within a single unified framework.

The DA dataset was generated using the same procedure as for the STG model (Section~\ref{sec:dataset_generation}). Due to a substantial fraction of silent populations in the DA parameter space, these were discarded, and the number of sampled configurations was set to yield approximately $25{,}000$ active populations. The same training protocol was used, with the LoRA-adapted model trained on this reduced dataset. The smaller dataset size is sufficient because LoRA leverages representations already learned by the STG backbone, requiring fewer examples to adapt to the new model.

\nolinenumbers

\bibliography{ref}